\documentclass[11pt,reqno]{article}
\usepackage[utf8]{inputenc}
\usepackage{amsmath,amssymb,amsthm,mathtools}
\usepackage{bm}
\usepackage[margin=1in]{geometry}
\usepackage{graphicx}
\usepackage{xcolor}
\usepackage[colorlinks=true,linkcolor=blue,citecolor=blue,urlcolor=blue]{hyperref}
\hypersetup{pdftitle={Perturbative Double-Scalar Hair on Planar Black Holes in AdS4-Einstein-Scalar Gravity},pdfauthor={Sangheon Yun}}
\usepackage{cite}
\usepackage[protrusion=true,expansion=true]{microtype}

\newcommand{\eps}{\epsilon}
\newcommand{\edens}{\varepsilon}
\newcommand{\pp}{\varphi_{+}}
\newcommand{\phim}{\varphi_{-}}
\newcommand{\rh}{r_{h}}
\newcommand{\Fhyp}{{}_{2}F_{1}}
\newcommand{\dd}{\mathrm{d}}
\newcommand{\Tr}{\operatorname{tr}}
\newcommand{\Order}[1]{O\!\left(#1\right)}
\newcommand{\csch}{\operatorname{csch}}
\newcommand{\Lop}{\mathcal{L}}
\newcommand{\Om}{\mathcal{O}_{-}}
\newcommand{\Oc}{\mathcal{O}_{\chi}}

\allowdisplaybreaks

\title{\bf Perturbative Double-Scalar Hair on Planar Black Holes\\ in AdS$_{4}$-Einstein-Scalar Gravity}

\author{Sangheon Yun\thanks{\texttt{sangheon.yun@gmail.com}}\\[6pt]
\small IndigoWave, Center for Quantum Spacetime, Sogang University,\\
\small 35 Baekbeom-ro, Mapo-gu, Seoul 04107, Republic of Korea}

\date{}

\begin{document}
\maketitle

\begin{abstract}
\noindent
We construct a two-scalar hairy planar black hole in four-dimensional Einstein gravity coupled to two scalars whose potential descends from the ABJM consistent truncation of eleven-dimensional supergravity.
Starting from the known single-scalar hairy black brane, we switch on the second scalar $\phim$ perturbatively in its dimensionless hair parameter $\eps$, which at fixed source is proportional to $\lambda_{-}T$, and solve the coupled Einstein-scalar system to order $\eps^{2}$.
The linearized $\phim$ profile is an exact hypergeometric function, independent of the $\pp$ hair, whose horizon-regular branch always carries a source, so there is no spontaneous $\phim$ hair.
Because the dual operator is irrelevant ($\Delta_{-}=4$), the second-order response develops a resonant logarithm: the condensate runs with a closed-form coefficient fixed by the cubic vertex.
The metric and $\pp$ back-reaction is obtained in closed form; for arbitrary hair it diagonalizes into two P\"oschl-Teller problems and locks the $\pp$ correction to the transverse metric correction with ratio $6/\sqrt7$.
On AdS-Schwarzschild the first law and holographic renormalization agree and give an analytic equation of state, $T/T_{0}-1=-\tfrac{5\sqrt3}{8\pi}\eps^{2}$ at fixed entropy and $c_{s}^{2}=\tfrac12-\tfrac{5\sqrt3}{8\pi}\eps^{2}$, with $\eta/s=1/4\pi$ and $\zeta/\eta=\tfrac14\eps^{2}$.
We then add the $\mathbb{Z}_{2}$-odd third scalar $\chi$ of the truncation, dual to $\Delta_{\chi}=5$.
On the hairy background its linear equation is a Heun equation with an apparent singularity, solved by a Clausen function.
At second order $\chi$ necessarily sources $\phim$, yet neither cross-coupling makes either condensate run, a mutual non-renormalization that follows from a one-step resonance identity and the relation $2\Delta_{\chi}-\Delta_{-}=2d$.
\end{abstract}

\vspace{2cm}

\newpage
\tableofcontents
\newpage

\section{Introduction}
\label{sec:intro}

Planar black-hole solutions of Einstein gravity coupled to scalar fields are the workhorses of applied holography: they realize finite-temperature states of strongly coupled field theories deformed by, or condensing, the operators dual to the bulk scalars \cite{HartnollLectures,HHH}.
When the scalar potential is not a free mass term but the exponential potential characteristic of a supergravity truncation, the resulting hairy branes capture genuine renormalization-group flows and can be analyzed in great detail.
A particularly clean example arises in the consistent truncation of eleven-dimensional supergravity relevant to ABJM theory \cite{ABJM,Gauntlett} (see also \cite{BakYun} for a related truncation), which retains two scalars $\pp$ and $\phim$ with a specific exponential potential.
The single-scalar sector, in which only $\pp$ is excited, admits an exact analytic hairy black brane \cite{Cadoni1,Cadoni2}, which was identified as a solution of the ABJM truncation in \cite{YunExact}.
There its holographic thermodynamics was also developed: with the source of the operator dual to $\pp$ held fixed, the plasma flows to the conformal ABJM fluid at low temperature, which is the regime underlying the boundary analysis below.

The natural next question, and the main subject of this paper, is what happens when the \emph{second} scalar $\phim$ is switched on.
Rather than attempt a fully nonlinear two-scalar solution, we treat the dimensionless hair parameter $\eps$ of $\phim$ as small and solve the coupled Einstein-scalar equations perturbatively, to order $\eps^{2}$.
This is the regime in which the back-reaction of $\phim$ on the geometry and on $\pp$ first appears, and it is analytically tractable throughout.

Two structural features make the problem unexpectedly transparent.
The first is that $\phim=0$ is a consistent truncation with the further property that the mixed second derivative $V_{,+-}$ of the potential vanishes on the truncation.
As a consequence there is no source for $\pp$ or the metric at first order in $\eps$, and the entire hair function of the background drops out of the $\phim$ sector.
The linearized $\phim$ equation is therefore identical for every member of the background family and reduces to a hypergeometric equation whose horizon-regular solution we write in closed form.
The second feature appears at second order: the operator $\Om$ is irrelevant in the three-dimensional boundary theory, with the integer dimension $\Delta_{-}=4=d+1$, so the quadratic source $\sim\phim^{2}$ resonates with the normalizable fall-off and produces a logarithm.
This is the holographic signature of a logarithmically running condensate, and we compute its coefficient in closed form.

The back-reaction is equally explicit.
On AdS-Schwarzschild the $g_{rr}$ correction is algebraic in the linear profile, and for arbitrary hair the coupled metric-$\pp$ response diagonalizes, in a constant basis, into two P\"oschl-Teller problems with elementary solutions.
On AdS-Schwarzschild the first law, applied at fixed physical source, then fixes the $\Order{\eps^{2}}$ thermodynamics analytically, and the resulting free energy agrees exactly with the holographically renormalized on-shell action.
Together with the horizon formula for the bulk viscosity, this yields closed-form expressions for the temperature shift, the sound speed and the bulk viscosity.
Throughout, a single point is essential: because $\Om$ is irrelevant, its source $\lambda_{-}$ is dimensionful, and the hair parameter is the effective coupling at the thermal scale, $\eps\propto\lambda_{-}T$. Thermodynamic and transport derivatives must therefore be taken at fixed $\lambda_{-}$, along which $\eps$ varies with the temperature, rather than at fixed $\eps$, which would compare thermal states of different theories.

\paragraph{A third scalar.}
The ABJM sector contains, besides the two dilatons, a further scalar $\chi$, odd under a $\mathbb{Z}_{2}$ reflection $\chi\to-\chi$, with a dilaton-dependent kinetic term and a potential even in $\chi$.
In the second part of the paper (Sections~\ref{sec:chi} and \ref{sec:chisecond}) we add $\chi$, with amplitude $\delta$, on the same footing as $\phim$, and ask how much of the structure above survives.
The answer is instructive.
The truncation $\chi=0$ is protected by symmetry rather than by a tuned cancellation, but the hair function no longer drops out: on the hairy background the linear $\chi$ equation becomes a Heun equation whose fourth singularity is \emph{apparent}, protected by the same identity that underlies the background, and whose regular solution is a Clausen function.
The equations of motion force the truncations to nest, so $\chi$ necessarily sources $\phim$ at second order.
Yet, unlike the $\phim$ self-coupling, neither cross-coupling makes either condensate run.
We trace this mutual non-renormalization to a one-step resonance identity that holds for arbitrary couplings, and show that it rests on the single dimension relation $2\Delta_{\chi}-\Delta_{-}=2d$.

\paragraph{Relation to earlier work.}
Exact and perturbative hairy planar AdS black holes have been studied extensively \cite{LuPopeWen,AnabalonHigherDim,YunG2,YunToda}, and the technical core of our linear analysis is standard: the horizon-regular linearized scalar on an AdS-Schwarzschild black brane is a Gauss hypergeometric function, and its leading back-reaction follows from radial quadratures fixed by horizon regularity, as in perturbative studies of non-conformal holographic plasmas \cite{KleinertProbst}.
The renormalization of sources for irrelevant operators, and the appearance of logarithms for integer dimensions, are known on general grounds \cite{vanReesA,vanReesB,SchwimmerTheisen,Broccoli}; scale-dependent multi-trace couplings in holography were developed in \cite{Papadimitriou,HeemskerkPolchinski}, and the systematics of anomalies in scalar three-point functions in \cite{BzowskiMcFaddenSkenderis}.
For the transport analysis we rely on the membrane-paradigm universality of $\eta/s$ \cite{BuchelLiu,IqbalLiu} and on the horizon formula for the bulk viscosity \cite{ElingOz}, whose validity has been tested in a number of non-conformal holographic plasmas \cite{BuchelEO,BuchelGursoyKiritsis} and which has been generalized to several scalars and vector fields \cite{HoyosKimOz}..
What is specific to the ABJM truncation, and new here, is:
(i) the exact decoupling $V_{,+-}|_{0}=0$, which renders the linearized $\phim$ equation identical for the entire hairy background family;
(ii) the closed-form coefficient of the induced condensate logarithm;
(iii) the constant-basis diagonalization of the general-hair back-reaction into two P\"oschl-Teller equations;
(iv) the analytic $\Order{\eps^{2}}$ equation of state and transport coefficients;
and (v) for the third scalar, the apparent-singularity Heun structure and the mutual non-renormalization of the two condensates, together with its general mechanism.

\paragraph{Outline.}
Section~\ref{sec:model} fixes the two-scalar model, and Section~\ref{sec:bg} reviews the single-scalar background.
Section~\ref{sec:scheme} sets up the perturbative expansion and solves the linearized $\phim$ equation exactly.
Section~\ref{sec:second} treats the second-order profile and derives the logarithmic running.
Section~\ref{sec:backreact} gives the gravitational back-reaction in closed form, first for the AdS-Schwarzschild background and then for arbitrary hair, and Section~\ref{sec:thermo} derives from it the equation of state and the transport coefficients.
Section~\ref{sec:chi} introduces the third scalar and solves its linear problem, and Section~\ref{sec:chisecond} analyzes its second-order interplay with $\phim$.
We conclude in Section~\ref{sec:disc}.
The appendices collect the potential data, the asymptotic recursion behind the logarithm, the second-order source induced by $\chi$, and the numerical checks.

\section{The two-scalar model}
\label{sec:model}

We work with four-dimensional Einstein gravity coupled to two canonically normalized scalars,
\begin{equation}
I=\frac{1}{16\pi G}\int \dd^{4}x\,\sqrt{-g}\left[\,R-\tfrac12(\partial\pp)^{2}-\tfrac12(\partial\phim)^{2}-V(\pp,\phim)\right],
\label{eq:action}
\end{equation}
with the potential
\begin{equation}
V = -\frac{12}{l^2}\,e^{-\frac{3\pp}{\sqrt7}+\frac{\phim}{\sqrt{21}}} + \frac{3}{2 l^2}\,e^{-\frac{3\pp}{\sqrt7}+\frac{8\phim}{\sqrt{21}}} + \frac{9}{2 l^2}\,e^{-\sqrt7\,\pp}.
\label{eq:V}
\end{equation}
This is the two-scalar potential of the ABJM truncation \cite{ABJM,Gauntlett,BakYun}; $l$ sets the asymptotic AdS$_{4}$ radius.
The equations of motion derived from \eqref{eq:action} are
\begin{equation}
G_{\mu\nu}=\tfrac12\,T_{\mu\nu},\qquad
T_{\mu\nu}=\sum_{a=\pm}\Big[\partial_{\mu}\varphi_{a}\,\partial_{\nu}\varphi_{a}-\frac{g_{\mu\nu}}{2}(\partial\varphi_{a})^{2}\Big]-g_{\mu\nu}V,
\label{eq:einstein}
\end{equation}
together with the two scalar equations $\Box\varphi_{a}=V_{,a}$, where $V_{,a}\equiv\partial V/\partial\varphi_{a}$.

\paragraph{Consistent truncation.}
Setting $\phim=0$ is consistent: \eqref{eq:V} gives
\begin{equation}
V_{,-}\big|_{\phim=0}=\frac{1}{l^{2}}\Big(-\tfrac{12}{\sqrt{21}}+\tfrac32\cdot\tfrac{8}{\sqrt{21}}\Big)e^{-3\pp/\sqrt7}=0 ,
\label{eq:notadpole}
\end{equation}
so the $\phim$ equation is solved identically by $\phim=0$ for any $\pp$.
On the truncation it is convenient to trade $\pp$ for $S\equiv e^{2\pp/\sqrt7}$, in terms of which the surviving single-scalar potential is
\begin{equation}
V\big|_{\phim=0}=-\frac{21}{2 l^2}\,S^{-3/2}+\frac{9}{2 l^2}\,S^{-7/2}.
\label{eq:Vtrunc}
\end{equation}
We abbreviate $\cdot|_{\phim=0}$ as $\cdot|_{0}$; expressions so evaluated remain functions of $S$, and $\cdot|_{0,S=1}$ denotes their further evaluation at the AdS vacuum $\pp=0$.
The data of the potential needed below are collected in Appendix~\ref{app:V}.
The single most important fact is that, in addition to \eqref{eq:notadpole}, the mixed derivative also vanishes on the truncation,
\begin{equation}
\boxed{\;V_{,+-}\big|_{0}=0\;}\,,
\label{eq:Vpm}
\end{equation}
which removes any $\Order{\eps}$ source for $\pp$ and the metric (Section~\ref{sec:scheme}).
The mass of $\phim$ around the truncation is
\begin{equation}
m_{-}^{2}\equiv V_{,--}\big|_{0}=\frac{4}{l^{2}}\,S^{-3/2}.
\label{eq:mass}
\end{equation}
At the AdS vacuum $S=1$, $m_{-}^{2}l^{2}=4$, and in AdS$_{4}$ (boundary dimension $d=3$) this corresponds to an operator $\Om$ of dimension
\begin{equation}
\Delta_{-}=\tfrac{d}{2}+\sqrt{\tfrac{d^{2}}{4}+m_{-}^{2}l^{2}}=\tfrac32+\tfrac52=4=d+1 ,
\end{equation}
i.e.\ an \emph{irrelevant} deformation of integer dimension.
This single number controls the qualitative physics of the second-order solution.
The vacuum $S=1$ is the conformal boundary of the AdS-Schwarzschild member of the background family only (Section~\ref{sec:bg}).

\section{The single-scalar hairy black brane}
\label{sec:bg}

We use the exact planar hairy solution of \cite{Cadoni1,Cadoni2}, in the ABJM-truncation parametrization of \cite{YunExact}, as our background.
In a domain-wall radial coordinate $\rho$,
\begin{eqnarray}
\dd s^2&=&-e^{2A_0(\rho)}\dd t^2+e^{2B_0(\rho)}\dd \rho^2 + e^{2H_0(\rho)}(\dd x_{1}^2+\dd x_{2}^2)\,,\nonumber \\
e^{A_0} &=& \frac{l\, e^{-2m \rho/3}}{3^{1/3}}\bigg( \frac{m^{4} \sinh^{3}(m \rho + \alpha)}{\sinh^{7} m \rho}\bigg)^{\!1/12},\qquad
e^{H_0} = \frac{e^{m \rho/3}}{3^{1/3}}\bigg( \frac{m^{4} \sinh^{3}(m \rho + \alpha)}{\sinh^{7} m \rho}\bigg)^{\!1/12}, \nonumber \\
e^{B_0} &=& \frac{l}{3}\bigg( \frac{m^{4} \sinh^{3}(m \rho + \alpha)}{\sinh^{7} m \rho}\bigg)^{\!1/4},\qquad
e^{\pp^{(0)}} = \bigg( \frac{\sinh(m \rho + \alpha)}{\sinh m \rho} \bigg)^{\!\sqrt{7}/2},
\label{eqs:background}
\end{eqnarray}
which satisfies $B_{0}=A_{0}+2H_{0}$.
In a radial coordinate $r$, related to $\rho$ by $r^{3}=ml^{2}[\coth(m\rho)+1]$, and after the constant rescalings $t\to3^{-1/3}l^{4/3}\,t$ and $x_{i}\to3^{-1/3}l^{-2/3}\,x_{i}$,
\begin{equation}
\dd s^{2}=-a^{2}\dd t^{2}+b^{2}\dd r^{2}+h^{2}(\dd x_{1}^{2}+\dd x_{2}^{2}),\qquad
a^{2}=f\,S^{1/2},\quad b^{2}=\frac{S^{3/2}}{f},\quad h^{2}=r^{2}S^{1/2},
\label{eq:bg}
\end{equation}
with $\pp^{(0)}=\tfrac{\sqrt7}{2}\ln S$ and
\begin{equation}
f(r)=\frac{r^{2}}{l^{2}}-\frac{2m}{r},\qquad
S(r)=\frac{\sinh\alpha}{m\,l^{2}}\,r^{3}+e^{-\alpha}.
\label{eq:fS}
\end{equation}
The two parameters are the mass parameter $m$ and the hair parameter $\alpha\ge0$; the value $\alpha=0$ gives $S\equiv1$, $\pp^{(0)}\equiv0$, and reduces \eqref{eq:bg} to the planar AdS-Schwarzschild$_{4}$ black brane.
One verifies directly that \eqref{eq:bg} solves \eqref{eq:einstein} and $\Box\pp=V_{,+}$ on the truncation \eqref{eq:Vtrunc}; the apparent obstruction from the $\sinh\alpha$ factors cancels by virtue of the identity
\begin{equation}
2e^{\alpha}\sinh\alpha=e^{2\alpha}-1 ,
\label{eq:identity}
\end{equation}
which will reappear, unexpectedly, in the $\chi$ sector (Section~\ref{sec:chiheun}).
The family enjoys the scaling symmetry $r\to\Omega r$, $m\to\Omega^{3}m$ at fixed $\alpha$, under which $f\to\Omega^{2}f$ and $S$ is invariant.

The horizon sits at $f(\rh)=0$, i.e.\ $\rh^{3}=2ml^{2}$, where $S(\rh)=e^{\alpha}$.
The Hawking temperature and entropy density are
\begin{equation}
T=\frac{1}{4\pi}\sqrt{(a^{2})'\,(b^{-2})'}\,\Big|_{\rh}=\frac{3\,\rh}{4\pi l^{2}}\,e^{-\alpha/2},\qquad
s=\frac{h^{2}(\rh)}{4G}=\frac{\rh^{2}\,e^{\alpha/2}}{4G},\qquad
Ts=\frac{3m}{8\pi G}.
\label{eq:bgthermo}
\end{equation}
It is convenient to use the dimensionless radial variable
\begin{equation}
x\equiv\frac{r^{3}}{\rh^{3}}\in[1,\infty),\qquad x=\big[1-e^{-2m\rho}\big]^{-1},
\label{eq:x}
\end{equation}
in terms of which $S$ is affine, $S=2\sinh\alpha\;x+e^{-\alpha}$.

\paragraph{Asymptotics.}
For $\alpha=0$ the geometry is asymptotically AdS$_{4}$ and the standard holographic dictionary applies.
For $\alpha>0$, however, $S\to\infty$ at large $r$, and $S\ge e^{\alpha}>1$ everywhere outside the horizon: the running $\pp$-hair deforms the ultraviolet into the hyperscaling-violating regime identified for this background in \cite{Cadoni2,CadoniSerra} (see \cite{Charmousis,GouterauxKiritsis} for the general framework), already visible in $h^{2}\sim r^{7/2}$.
For the $\pp$-hair itself, a holographic dictionary adapted to these asymptotics was constructed in \cite{YunExact} by the Hamilton--Jacobi method \cite{KST}, under explicitly stated working assumptions.
In the resulting Dirichlet theory the source of the operator dual to $\pp$ is held fixed and defines a scale $\Lambda$, and $\alpha$ is not an independent parameter but determines $T/\Lambda$; on the thermodynamically stable branch, $\alpha\simeq\tfrac12(4\pi T/3\Lambda)^{3}$ at low temperature.
The AdS-Schwarzschild member $\alpha=0$ is the ABJM fixed point itself, reached as $T/\Lambda\to0$; it is also the $\hat k=0$ member of the scale-invariant multi-trace theory of \cite{YunExact}, which holds $\alpha$ fixed instead.
What we do not attempt is the corresponding dictionary for the new sources $\lambda_{-}$ and $\lambda_{\chi}$ switched on in this paper, which at $\alpha>0$ would require extending that renormalization.
Accordingly, statements about the new sources, condensates, operator dimensions and the thermodynamics of the deformations are made on the AdS-Schwarzschild member $\alpha=0$, i.e.\ for deformations of the ABJM fixed point; within the Dirichlet theory these are the leading terms at $T\ll\Lambda$, the coupling to the $\pp$ flow entering at relative order $(T/\Lambda)^{3}$.
For $\alpha>0$ we quote the coefficients of the independent radial fall-offs and the horizon data, which are well defined.

\section{Perturbative scheme and the linearized scalar}
\label{sec:scheme}

We switch on $\phim$ with amplitude $\eps$ and expand all fields,
\begin{equation}
\phim=\eps\,\psi_{1}+\eps^{2}\psi_{2}+\cdots,\qquad
\pp=\pp^{(0)}+\eps^{2}\,\Phi_{2}+\cdots,\qquad
g_{\mu\nu}=g^{(0)}_{\mu\nu}+\eps^{2}\,g^{(2)}_{\mu\nu}+\cdots.
\label{eq:expand}
\end{equation}
Because $V_{,+-}|_{0}=0$ \eqref{eq:Vpm}, there are no $\Order{\eps}$ corrections to $\pp$ or to the metric: the leading back-reaction is quadratic.
At $\Order{\eps}$ the only nontrivial equation is the linearized $\phim$ equation on the fixed background \eqref{eq:bg},
\begin{equation}
\Box_{0}\,\psi_{1}=m_{-}^{2}\,\psi_{1},
\label{eq:lin}
\end{equation}
where $\Box_{0}$ is the d'Alembertian of $g^{(0)}_{\mu\nu}$.
A remarkable simplification occurs.
Since $\sqrt{-g_{0}}=S^{3/2}r^{2}$, $\sqrt{-g_{0}}\,g_{0}^{rr}=fr^{2}$ and $m_{-}^{2}\propto S^{-3/2}$, the hair function $S$ cancels completely, and \eqref{eq:lin} is \emph{independent of $\alpha$}:
\begin{equation}
\frac{\dd}{\dd r}\!\Big[(r^{4}-2ml^{2}r)\,\psi_{1}'\Big]=4\,r^{2}\,\psi_{1}.
\label{eq:linr}
\end{equation}
In the variable $x$ of \eqref{eq:x}, which removes $m$ as well, \eqref{eq:linr} becomes the Gauss hypergeometric equation
\begin{equation}
\Lop[\psi_{1}]=0,\qquad
\Lop\equiv x(x-1)\,\partial_{x}^{2}+(2x-1)\,\partial_{x}-\tfrac49 ,
\label{eq:hyp}
\end{equation}
with indices $(a,b;c)=(\tfrac43,-\tfrac13;1)$.
The solution regular at the horizon $x=1$, normalized to $\psi_{1}(\rh)=1$, is
\begin{equation}
\boxed{\;\psi_{1}(x)=\Fhyp\!\Big(\tfrac43,-\tfrac13;1;\,1-x\Big)\;}
=1+\tfrac49\,t-\tfrac{14}{81}\,t^{2}+\tfrac{700}{6561}\,t^{3}-\tfrac{4550}{59049}\,t^{4}+\cdots,\qquad t\equiv x-1 .
\label{eq:psi1}
\end{equation}

\paragraph{Boundary data.}
Using the connection formulas for the hypergeometric function \cite{DLMF}, \eqref{eq:psi1} can be written, for all $x>1$, as the exact superposition of two branches,
\begin{equation}
\psi_{1}=C_{s}\,(x-1)^{1/3}\,\Fhyp\!\Big(-\tfrac13,-\tfrac13;-\tfrac23;\tfrac{1}{1-x}\Big)
+C_{v}\,(x-1)^{-4/3}\,\Fhyp\!\Big(\tfrac43,\tfrac43;\tfrac83;\tfrac{1}{1-x}\Big),
\label{eq:psi1bndry}
\end{equation}
with constant coefficients
\begin{equation}
C_{s}=\frac{6\,\Gamma(\tfrac23)}{\Gamma(\tfrac13)^{2}}\approx1.1321,\qquad
C_{v}=\frac{\Gamma(\tfrac13)}{10\,\Gamma(\tfrac23)^{2}}\approx0.1461,\qquad
C_{s}C_{v}=\frac{3\sqrt3}{10\pi}.
\label{eq:CsCv}
\end{equation}
Since $(x-1)^{1/3}\sim r$ and $(x-1)^{-4/3}\sim r^{-4}$, on AdS-Schwarzschild $C_{s}$ multiplies the source of $\Om$ (fall-off $r^{\Delta_{-}-d}=r$) and $C_{v}$ its response (fall-off $r^{-\Delta_{-}}=r^{-4}$).
In the Fefferman-Graham coordinate $z\simeq l^{2}/r$,\footnote{Since $l^{2}\dd z^{2}/z^{2}=l^{2}\dd r^{2}/r^{2}$ for $z=c/r$ with any constant $c$, the normalization of $z$ is a choice of boundary units. With $z\simeq l^{2}/r$ the boundary metric of \eqref{eq:bg} at $\alpha=0$ is $-\dd t^{2}+l^{2}\dd\vec x^{\,2}$, so that $t$ is the physical boundary time, in terms of which $T$ in \eqref{eq:bgthermo} is measured, while all densities, such as $s=\rh^{2}/4G$, are quoted per unit coordinate area $\dd x_{1}\dd x_{2}$. The choice $z\simeq l/r$ would rescale $\lambda_{-}$ and $T$ by inverse constant factors and leave the dimensionless combination $\lambda_{-}T$ in \eqref{eq:lambdaminus} unchanged.} $\phim\simeq\lambda_{-}z^{-1}+\cdots+\mathcal{V}_{-}z^{4}$ with
\begin{equation}
\lambda_{-}=\eps\,C_{s}\,\frac{l^{2}}{\rh}+\Order{\eps^{2}},\qquad
\mathcal{V}_{-}=\eps\,C_{v}\Big(\frac{\rh}{l^{2}}\Big)^{4}+\Order{\eps^{2}},
\qquad\text{i.e.}\qquad
\eps=\frac{4\pi\,\lambda_{-}T}{3\,C_{s}}+\Order{\lambda_{-}^{2}T^{2}} .
\label{eq:lambdaminus}
\end{equation}
The dimensionless hair parameter $\eps$ is thus the effective coupling $\lambda_{-}T$ of the irrelevant deformation at the thermal scale: at fixed physical source, $\eps$ grows linearly with the temperature.
For $\alpha>0$ the profile \eqref{eq:psi1}, the coefficients \eqref{eq:CsCv} and the radial fall-offs $r$, $r^{-4}$ are unchanged, owing to the cancellation of $S$; only their holographic interpretation awaits the renormalization noted in Section~\ref{sec:bg}.

\paragraph{No spontaneous $\phim$ hair.}
For every background \eqref{eq:bg}, the horizon-regular linearized hair \eqref{eq:psi1} contains the growing branch, $C_{s}\neq0$.
Since $\sqrt{-g_{0}}\,g_{0}^{rr}=fr^{2}$ for every $\alpha$, the $r^{1}$ branch has divergent kinetic norm and the $r^{-4}$ branch a finite one, exactly as in AdS; hence there is no choice of $(\alpha,m)$ for which the regular hair is normalizable, and on AdS-Schwarzschild $C_{s}\neq0$ is a nonzero source.
Consequently $\phim$ can be turned on only as an explicit, irrelevant deformation, accompanied by an induced condensate.

\section{Second order: the profile \texorpdfstring{$\psi_{2}$}{psi2} and logarithmic running}
\label{sec:second}

At $\Order{\eps^{2}}$ the $\phim$ equation acquires the source $\tfrac12V_{,---}|_{0}\,\psi_{1}^{2}$ (Appendix~\ref{app:V}).
Because the metric and $\pp$ corrections enter the $\phim$ equation only at $\Order{\eps^{3}}$, and $V_{,---}|_{0}\propto S^{-3/2}$ so that the hair again cancels, the equation for $\psi_{2}$ closes on the background for every $\alpha$:
\begin{equation}
\frac{\dd}{\dd r}\!\Big[(r^{4}-2ml^{2}r)\,\psi_{2}'\Big]-4r^{2}\psi_{2}
=\frac{6\sqrt{21}}{7}\,r^{2}\,\psi_{1}^{2},
\qquad\text{i.e.}\qquad
\Lop[\psi_{2}]=\frac{2}{\sqrt{21}}\,\psi_{1}^{2}.
\label{eq:psi2x}
\end{equation}
Like $\psi_{1}$, the function $\psi_{2}$ is independent of $\alpha$ and $m$.

\paragraph{An identity and the algebraic part.}
Acting with $\Lop$ on $\psi_{1}^{2}$ and using $\Lop[\psi_{1}]=0$ gives
\begin{equation}
\Lop[\psi_{1}^{2}]=\tfrac49\,\psi_{1}^{2}+2\,x(x-1)\,\dot{\psi}_{1}^{2},
\label{eq:Lpsi1sq}
\end{equation}
where dots denote $\dd/\dd x$.
This isolates the algebraic part of the response:
\begin{equation}
\psi_{2}=\frac{3\sqrt{21}}{14}\,\psi_{1}^{2}+\hat\psi_{2},\qquad
\Lop[\hat\psi_{2}]=-\frac{3\sqrt{21}}{7}\,x(x-1)\,\dot{\psi}_{1}^{2}.
\label{eq:psi2decomp}
\end{equation}
The remainder $\hat\psi_{2}$ is fixed by quadrature.
The second homogeneous solution, obtained by reduction of order,
\begin{equation}
\widetilde{\psi}(x)=\psi_{1}(x)\int^{x}\frac{\dd \hat{x}}{\hat{x}(\hat{x}-1)\,\psi_{1}(\hat{x})^{2}} ,
\label{eq:psitilde}
\end{equation}
is logarithmically singular at the horizon.
With the Wronskian normalized to $x(x-1)\,W[\psi_{1},\widetilde\psi]=1$, variation of parameters yields
\begin{equation}
\psi_{2}^{\mathrm{p}}(x)=\frac{2}{\sqrt{21}}\left[\widetilde\psi(x)\int^{x}\!\psi_{1}^{3}\,\dd \hat{x}-\psi_{1}(x)\int^{x}\!\widetilde\psi\,\psi_{1}^{2}\,\dd \hat{x} \right].
\label{eq:psi2p}
\end{equation}
The horizon-regular solution is $\psi_{2}=\psi_{2}^{\mathrm{p}}+d_{0}\,\psi_{1}$, with the lower limits chosen to remove the logarithm at $x=1$.
Normalized by $\psi_{2}(\rh)=0$, so that $\eps$ remains the horizon value of $\phim$, its Frobenius series has coefficients that are rational multiples of $\sqrt{21}$,
\begin{equation}
\psi_{2}=\frac{2\sqrt{21}}{21}\,t-\frac{\sqrt{21}}{63}\,t^{2}+\frac{2\sqrt{21}}{243}\,t^{3}-\frac{110\sqrt{21}}{19683}\,t^{4}+\cdots .
\label{eq:psi2reg}
\end{equation}

\paragraph{Boundary asymptotics: resonance and the logarithm.}
The boundary expansion of $\psi_{2}$ is governed by the indicial polynomial of $\Lop$, which factorizes as $P(q)=(q+\tfrac43)(q-\tfrac13)$, with zeros at the source exponent $q=\tfrac13$ and the response exponent $q=-\tfrac43$.
The source $\psi_{1}^{2}$ drives a tower of powers $t^{2/3-n}$ descending in integer steps; the $n=2$ member coincides with the normalizable fall-off $t^{-4/3}$, where $P$ vanishes.
The forcing is therefore resonant, two steps below its leading power, and a logarithm is generated.
Solving the asymptotic recursion (Appendix~\ref{app:recursion}) gives
\begin{equation}
\boxed{\;
\begin{aligned}
\psi_{2}&=\frac{\sqrt{21}}{7}C_{s}^{2}\,t^{2/3}
+A_{s}\,t^{1/3}
+\frac{\sqrt{21}}{21}C_{s}^{2}\,t^{-1/3}
-\frac{9\,C_{s}C_{v}}{\sqrt{21}}\,t^{-1}
+\Big[a_{-4/3}+\mathsf{A}\,C_{s}^{2}\,\ln t\Big]t^{-4/3}+\cdots,\\
\mathsf{A}&=\frac{\sqrt{21}}{70}
\end{aligned}\;}
\label{eq:psi2bndry}
\end{equation}
where $A_{s}$ multiplies the complete source branch.
The leading term $\sim r^{2}$ is the standard ultraviolet growth induced by an irrelevant deformation: the expansion is controlled for $r\ll\rh/\eps$ and breaks down beyond.
The coefficients of the forced tower are fixed by the source, while $A_{s}$ and $a_{-4/3}$ are shifted by the homogeneous admixture $d_{0}\psi_{1}$; for the normalization \eqref{eq:psi2reg} we find numerically $A_{s}\approx-0.590$ and $a_{-4/3}\approx0.207$.
The coefficient of the logarithm, however, is fixed, and since the forcing is linear in $V_{,---}$ it is proportional to the cubic vertex,
\begin{equation}
\mathsf{A}=\frac{\sqrt{21}}{70}=\frac{l^{2}}{120}\,V_{,---}\big|_{0,S=1},
\label{eq:anomcubic}
\end{equation}
the resonance supplying only the rational factor $1/120$.

\paragraph{Logarithmic running of the condensate.}
On AdS-Schwarzschild, $t\simeq x=(r/\rh)^{3}=(l^{2}/z\rh)^{3}$, so that $t^{-4/3}\ln t=-3\,(z\rh/l^{2})^{4}\ln(z\rh/l^{2})$.
The $z^{4}\ln z$ term is removed by a local counterterm at the cutoff, cubic in $\phim$, whose finite part is not fixed by the bulk data \cite{vanReesA}; it introduces a renormalization scale $\mu$.
The coefficient of the normalizable mode, which determines $\langle\Om\rangle$ up to the conventional normalization factor and scheme-dependent constants, then reads
\begin{equation}
\mathcal{V}_{-}=\Big(\frac{\rh}{l^{2}}\Big)^{4}\Big\{\eps\,C_{v}+\eps^{2}\Big[a_{-4/3}'+3\mathsf{A}\,C_{s}^{2}\ln\frac{\mu\, l^{2}}{\rh}\Big]+\cdots\Big\},
\label{eq:OmVEV}
\end{equation}
with $a_{-4/3}'$ a scheme-dependent constant, and
\begin{equation}
\mu\,\frac{\partial\mathcal{V}_{-}}{\partial\mu}=3\mathsf{A}\,C_{s}^{2}\,\eps^{2}\Big(\frac{\rh}{l^{2}}\Big)^{4}
=3\mathsf{A}\,\lambda_{-}^{2}\Big(\frac{\rh}{l^{2}}\Big)^{6},
\qquad 3\mathsf{A}=\frac{3\sqrt{21}}{70}\approx0.196 ,
\label{eq:beta}
\end{equation}
using \eqref{eq:lambdaminus}; the factor $3=d$ reflects $t\simeq(r/\rh)^{3}$.
The slope is independent of the scheme and of the normalization of $\psi_{2}$.

\paragraph{Physical content.}
Each step down the tower costs one power of $x^{-1}\propto m$, i.e.\ of the energy density, so the logarithm \eqref{eq:beta} multiplies $\lambda_{-}^{2}(\rh/l^{2})^{6}\propto\lambda_{-}^{2}\langle T_{tt}\rangle^{2}$.
It is therefore not a local anomaly functional of the sources and the boundary metric, which would vanish for constant sources in flat space, but the state-dependent kind of running that irrelevant deformations generate \cite{vanReesA,vanReesB}.
One natural reading, in the spirit of multi-trace renormalization \cite{Papadimitriou,HeemskerkPolchinski,vanReesA}, is the following.
Since $\lambda_{-}$ has dimension $d-\Delta_{-}=-1$, the double-trace coupling $\lambda_{-}^{3}\int[T_{\mu\nu}T^{\mu\nu}]$ is marginal, and its coefficient $c(\mu)$ can run logarithmically; at large $N$, $\langle[T_{\mu\nu}T^{\mu\nu}]\rangle\propto T^{6}$, so that $-\partial_{\lambda_{-}}$ of this term reproduces the structure of \eqref{eq:beta}.
We have not derived this identification from the bulk counterterm, and the normalization of $c(\mu)$ would provide an independent field-theory check of $\mathsf{A}$.
Consistently, the logarithm first enters the free energy at $\Order{\eps^{3}}$, through $\partial F/\partial\lambda_{-}=-\langle\Om\rangle$.

\paragraph{Relation to the three-point function.}
The coefficient \eqref{eq:anomcubic} originates entirely in the cubic vertex $V_{,---}$, which also generates the AdS$_{4}$ Witten diagram for $\langle\Om\Om\Om\rangle$ and hence sets the $\Om\Om\Om$ OPE coefficient.
The thermal logarithm must nonetheless be distinguished from the vacuum anomaly of that correlator: by the analysis of \cite{BzowskiMcFaddenSkenderis}, the latter arises because $3\Delta_{-}-2d=6=2k$ with $k=3$, and multiplies six derivatives of the sources, so that it is invisible in a homogeneous state.
The two share their origin in $V_{,---}$, but their normalizations are independent data, and an independent field-theory determination of either would be a useful check.

\paragraph{What is observable.}
Because the slope is fixed while the constant $a_{-4/3}'$ is not, the unambiguous content is a difference.
At fixed physical source $\lambda_{-}$, the $\Order{\lambda_{-}^{2}}$ part of the condensate satisfies
\begin{equation}
T\frac{\dd}{\dd T}\bigg[\frac{\mathcal{V}_{-}^{(2)}}{\lambda_{-}^{2}(\rh/l^{2})^{6}}\bigg]=-3\mathsf{A},
\end{equation}
so that comparing two temperatures isolates $\mathsf{A}$, the scheme-dependent constant cancelling in the difference.

\section{Gravitational back-reaction in closed form}
\label{sec:backreact}

At $\Order{\eps^{2}}$ the metric functions obey first-order radial equations whose source is quadratic in the linear data, and which integrate to quadratures fixed by horizon regularity, as is standard for second-order holographic back-reaction \cite{KleinertProbst}.
Two features are special to the present problem: on AdS-Schwarzschild the $g_{rr}$ correction is \emph{algebraic} in the linear profile, and for arbitrary hair (Section~\ref{sec:generalalpha}) the two gravitational unknowns and the $\pp$ correction assemble into a system with \emph{constant} coefficients in a suitable basis, which diagonalizes exactly into two decoupled P\"oschl-Teller problems with elementary solutions.

\subsection{AdS-Schwarzschild background (\texorpdfstring{$\alpha=0$}{alpha=0})}
\label{sec:alpha0}

For $\alpha=0$ the background scalar vanishes, $\pp^{(0)}=0$, and the black hole is the AdS-Schwarzschild$_{4}$ brane with $a^{2}=b^{-2}=f$ and $h^{2}=r^{2}$.
Two facts decouple the second-order problem.
First, the coupling between the metric and the $\pp$ perturbation is proportional to $\pp^{(0)\prime\prime}=0$ and so vanishes.
Second, since $V_{,+}|_{0,S=1}=0$ (the AdS extremum), the $\pp$ correction $\Phi_{2}$ does not source the metric at $\Order{\eps^{2}}$.
The three sectors, $\psi_{2}$, $\Phi_{2}$ and the metric, are thus independent.
We parametrize the corrected metric in areal gauge,
\begin{equation}
\dd s^{2}=-f\big(1+\eps^{2}u\big)\dd t^{2}+\frac{\dd r^{2}}{f\big(1+\eps^{2}v\big)}+r^{2}\big(\dd x_{1}^{2}+\dd x_{2}^{2}\big).
\label{eq:arealmetric}
\end{equation}
Since the result will be reused for the third scalar in Section~\ref{sec:chigeom}, we write the $(tt)$ and $(rr)$ Einstein equations for a static scalar $\varphi$ with kinetic weight $k$ and potential curvature $M^{2}$ (for $\phim$, $\varphi=\psi_{1}$, $k=1$, $M^{2}l^{2}=4$):
\begin{align}
\frac{\dd}{\dd r}\Big[(r^{3}-\rh^{3})\,v\Big]&=-\frac{k}{4}\,r\,(r^{3}-\rh^{3})\,\varphi'^{2}-\frac{M^{2}l^{2}}{4}\,r^{2}\varphi^{2},
\label{eq:v}\\
u'&=\frac{k}{4}\,r\,\varphi'^{2}-\frac{r^{2}\big(M^{2}l^{2}\varphi^{2}+12\,v\big)}{4\,(r^{3}-\rh^{3})},
\label{eq:u}
\end{align}
while the $(x_{i}x_{i})$ equation is satisfied identically, as required by the Bianchi identity.

\paragraph{Closed form for $v$.}
The $(tt)$ equation involves only $v$ and can be carried out in closed form.
If $\varphi$ solves the linear equation, $k\Box\varphi=M^{2}\varphi$, then in the variable $x$
\begin{equation}
\frac{\dd}{\dd x}\Big[x(x-1)\,\varphi\dot\varphi\Big]=x(x-1)\,\dot\varphi^{2}+c\,\varphi^{2},
\qquad c=\frac{M^{2}l^{2}}{9k},
\label{eq:vtotalderiv}
\end{equation}
with $c=\tfrac49$ for $\phim$.
Rewriting \eqref{eq:v} in terms of $x$ and using \eqref{eq:vtotalderiv}, the two $\varphi^{2}$ contributions cancel identically, for any minimally coupled scalar, and \eqref{eq:v} integrates in a single step to
\begin{equation}
\boxed{\;v(x)=-\frac{3k}{4}\,x\,\varphi(x)\,\dot\varphi(x)\;}\,,
\qquad\text{for $\phim$:}\quad v=-\tfrac34\,x\,\psi_{1}\dot\psi_{1},
\label{eq:vclosed}
\end{equation}
valid for all $x\ge1$ with no residual integral, the integration constant being fixed by regularity at the horizon.
The horizon value follows immediately from $\psi_{1}(1)=1$ and $\dot\psi_{1}(1)=\tfrac49$:
\begin{equation}
\boxed{\;v(\rh)=-\tfrac34\cdot1\cdot\tfrac49=-\tfrac13\;}\,,
\qquad\text{and in general}\quad v(\rh)=-\frac{M^{2}l^{2}}{12}.
\label{eq:vh}
\end{equation}
At large $x$, $v\simeq-\tfrac14C_{s}^{2}x^{2/3}$, so the back-reaction becomes of order one at $r\sim\rh/\eps$, the scale at which the dimensionless strength of the deformation reaches unity; the perturbative geometry is reliable well inside this radius.

\paragraph{The function $u$.}
The $(rr)$ equation \eqref{eq:u} then determines $u$ by a single further quadrature, now with $v$ known explicitly.
At large $r$ the leading growth of $u'$ cancels,
\begin{equation}
u'(r)\simeq\frac{3C_{s}^{2}\,\rh}{8r^{2}}\qquad(r\to\infty),
\end{equation}
so $u(\infty)$ is finite and the boundary time can be canonically normalized by $u(\infty)=0$.
Numerical integration then gives $u(\rh)=-0.355827785943907\ldots$, in agreement to fifteen digits with the closed form $u(\rh)=\tfrac13-\tfrac{5\sqrt3}{4\pi}$ that we derive from the first law in Section~\ref{sec:eos}.
We have not found a closed form for the full profile $u(x)$: an ansatz $u=a(x)\psi_{1}^{2}+b(x)\psi_{1}\dot\psi_{1}+c(x)\dot\psi_{1}^{2}$ with rational coefficients, modeled on \eqref{eq:vtotalderiv}, admits no solution.
Since $u$ and $v$ depend on $r$ only through $x$, the resulting temperature shift is independent of $\rh$ and $l$.

\paragraph{The dilaton correction.}
The $\pp$ correction obeys
\begin{equation}
\frac{\dd}{\dd r}\Big[(r^{4}-2ml^{2}r)\,\Phi_{2}'\Big]-18\,r^{2}\,\Phi_{2}=-\frac{6}{\sqrt7}\,r^{2}\psi_{1}^{2},
\label{eq:Phi2}
\end{equation}
i.e.\ a hypergeometric equation with indices $(2,-1;1)$, corresponding to the $\Delta=6$ fluctuation of $\pp$ around AdS.
Both homogeneous solutions are elementary,
\begin{equation}
\Phi^{\mathrm{reg}}=2x-1,\qquad
\widetilde\Phi=(2x-1)\ln\frac{x-1}{x}+2,
\label{eq:Phihom}
\end{equation}
with $x(x-1)\,W[\Phi^{\mathrm{reg}},\widetilde\Phi]=1$, and the particular solution follows from \eqref{eq:psi2p} with the obvious substitutions.
Thus at $\alpha=0$ the dilaton vanishes only at zeroth order: it is necessarily induced at $\Order{\eps^{2}}$, in accordance with the nested truncation of Section~\ref{sec:chimodel}, but it requires no new source, since the forcing $\psi_{1}^{2}$, whose large-$x$ tower $x^{2/3-n}$ never reaches the fall-offs $x^{1}$ and $x^{-2}$ of the homogeneous solutions \eqref{eq:Phihom}, is non-resonant, and the coefficient of the source mode $2x-1$ can be set to zero. The parameters $\alpha$ and $\eps$ are therefore independent, and Section~\ref{sec:generalalpha} shows that this remains true for arbitrary hair.

\subsection{Arbitrary hair (general \texorpdfstring{$\alpha$}{alpha})}
\label{sec:generalalpha}

For $\alpha\neq0$ the metric and $\pp$ corrections couple.
We write the perturbed metric in the domain-wall form of \eqref{eqs:background},
\begin{equation}
\dd s^2 = -e^{2A_0+2\eps^{2}A_{2}}\dd t^2 + e^{2B_0+2\eps^{2}B_{2}}\dd \rho^2 + e^{2H_0+2\eps^{2}H_{2}}(\dd x_{1}^2+\dd x_{2}^2),
\qquad \pp=\pp^{(0)}+\eps^{2}\Phi_{2},
\end{equation}
and preserve the gauge $B=A+2H$ of the background, $B_{2}=A_{2}+2H_{2}$, in which $\sqrt{-g}=e^{2B}$ and the dynamical equations read $A''=H''=-\tfrac12e^{2B}V$ and $\pp''=e^{2B}V_{,+}$.
These imply that $A_{2}-H_{2}$ is linear in $\rho$; this is a homogeneous mode, which can be reinstated at the end and which we set to zero, $A_{2}=H_{2}$.
The coupled system for $\bm{w}=(H_{2},\Phi_{2})^{T}$ then takes the form
\begin{equation}
\bm{w}''=\bm{M}(\rho)\,\bm{w}+\bm{S}(\rho),\qquad
\bm{M}=\begin{pmatrix}6H_{0}''&-\tfrac12\pp^{(0)\prime\prime}\\[3pt]6\,\pp^{(0)\prime\prime}&W_{0}V_{,++}|_{0}\end{pmatrix},
\label{eq:coupled}
\end{equation}
with $W_{0}=e^{2B_{0}}$ and source $\bm{S}=\big(-\tfrac14 W_{0}m_{-}^{2}\,\psi_{1}^{2},\ \tfrac12 W_{0}V_{,+--}|_{0}\,\psi_{1}^{2}\big)^{T}$, supplemented by the first-order $(\rho\rho)$ constraint.
In this gauge the background reads $\pp^{(0)}=\tfrac{\sqrt7}{2}\ln[\sinh(m\rho+\alpha)/\sinh(m\rho)]$.

The key structural fact is that $\bm{M}$ factorizes into constant matrices times scalar functions.
Writing $\mathcal{A}=\csch^{2}(m\rho+\alpha)$ and $\mathcal{B}=\csch^{2}(m\rho)$, one finds
\begin{equation}
\boxed{\;\bm{M}(\rho)=m^{2}(\mathcal{A}-\mathcal{B})\,\bm{M}_{A}+2m^{2}\mathcal{B}\,I\;},\qquad
\bm{M}_{A}=\begin{pmatrix}-\tfrac32&\tfrac{\sqrt7}{4}\\[3pt]-3\sqrt7&\tfrac72\end{pmatrix}.
\label{eq:Mdecomp}
\end{equation}
Since $\bm{M}_{A}$ is constant, its eigenvectors are $\rho$-independent and the system diagonalizes in a fixed basis.
With $\Tr\bm{M}_{A}=2$ and $\det\bm{M}_{A}=0$, the eigenvalues are $\kappa\in\{2,0\}$, with eigenvectors
\begin{equation}
\bm{e}_{2}=\begin{pmatrix}1\\2\sqrt7\end{pmatrix}\ (\kappa=2),\qquad
\bm{e}_{0}=\begin{pmatrix}\sqrt7\\6\end{pmatrix}\ (\kappa=0).
\end{equation}
Writing $\bm{w}=w_{2}\,\bm{e}_{2}+w_{0}\,\bm{e}_{0}$, the homogeneous problem reduces to two decoupled scalar equations of P\"oschl-Teller type,
\begin{equation}
w_{2}''=\frac{2m^{2}}{\sinh^{2}(m\rho+\alpha)}\,w_{2},\qquad
w_{0}''=\frac{2m^{2}}{\sinh^{2}(m\rho)}\,w_{0},
\end{equation}
each of the form $y''=2m^{2}\sinh^{-2}(\xi)\,y$ with $\xi=m(\rho-\rho_{0})$, and each with elementary solutions
\begin{equation}
y_{1}=\coth\xi,\qquad y_{2}=\xi\coth\xi-1,\qquad W[y_{1},y_{2}]=m .
\end{equation}
The four solutions of the homogeneous system are thus all elementary; their $4\times4$ Wronskian equals $(\det[\bm{e}_{2},\bm{e}_{0}])^{2}m^{2}=64m^{2}$ up to sign.
Since \eqref{eqs:background} solves the field equations for every $(m,\alpha)$, the parameter derivatives of the background solve the linearized equations.
The hair modulus, $\partial_{\alpha}(\text{background})=\tfrac14\coth(m\rho+\alpha)\,\bm{e}_{2}$, preserves $A_{2}=H_{2}$ and is therefore a solution of the homogeneous system \eqref{eq:coupled}.
The derivative $\partial_{m}(\text{background})=\tfrac{\rho}{m}\partial_{\rho}(\text{background})+\tfrac{1}{3m}(1,0)^{T}$, by contrast, carries $A_{2}-H_{2}=-\rho$ and so lies in the linear mode set to zero above; by the scaling symmetry of Section~\ref{sec:bg} it is pure gauge, a radial diffeomorphism $\rho\to\rho(1+\delta m/m)$ combined with a constant rescaling of $t$ and $x_{i}$.

Projecting the source with the left eigenvectors $\bm{f}_{2}=(-\tfrac34,\tfrac{\sqrt7}{8})$ and $\bm{f}_{0}=(\tfrac{\sqrt7}{4},-\tfrac18)$, which form the basis dual to $\{\bm{e}_{2},\bm{e}_{0}\}$, produces a striking simplification:
\begin{equation}
\boxed{\;\sigma_{2}\equiv\bm{f}_{2}\!\cdot\!\bm{S}=0\;},\qquad
\sigma_{0}\equiv\bm{f}_{0}\!\cdot\!\bm{S}=-\frac{4\sqrt7\,m^{2}}{63}\,x(x-1)\,\psi_{1}^{2}.
\label{eq:sigma}
\end{equation}
The $\kappa=2$ direction, which is the $\partial_{\alpha}$ modulus, is not sourced: switching on $\phim$ does not shift the hair parameter at this order.
The entire physical response lies along $\bm{e}_{0}$, so that
\begin{equation}
\boxed{\;H_{2}=\sqrt7\,w_{0},\qquad \Phi_{2}=6\,w_{0}\quad\Longrightarrow\quad \Phi_{2}=\frac{6}{\sqrt7}\,H_{2}\;},
\end{equation}
up to the homogeneous $\bm{e}_{2}$ mode fixed by the boundary conditions: the $\pp$ correction is locked to the transverse metric correction in the constant ratio $6/\sqrt7$.

It remains to give $w_{0}$.
In the variable $x$ one has $\coth(m\rho)=2x-1$ and $m\rho=\tfrac12\ln\frac{x}{x-1}$, so the $\kappa=0$ homogeneous solutions coincide with those of the $\alpha=0$ problem,
\begin{equation}
y_{1}^{(0)}=2x-1=\Phi^{\mathrm{reg}},\qquad y_{2}^{(0)}=\frac{2x-1}{2}\ln\frac{x}{x-1}-1=-\frac{1}{2}\widetilde\Phi .
\end{equation}
Variation of parameters, with $\dd \rho=-\dd x/[2mx(x-1)]$, gives the closed-form quadrature
\begin{equation}
\boxed{\;
w_{0}(x)=\frac{2\sqrt7}{63}\left[y_{2}^{(0)}(x)\!\int^{x}\!(2\hat{x}-1)\,\psi_{1}^{2}\,\dd \hat{x}-(2x-1)\!\int^{x}\!y_{2}^{(0)}\,\psi_{1}^{2}\,\dd \hat{x} \right]
+b_{1}(2x-1)+b_{2}\,y_{2}^{(0)}(x)\;},
\label{eq:zeta0}
\end{equation}
with $b_{1},b_{2}$ fixed by horizon regularity (removal of the logarithmic mode as $x\to1$) and by the $(\rho\rho)$ constraint.
This completes the construction of the double-scalar hairy brane to $\Order{\eps^{2}}$ for arbitrary hair.
\section{Thermodynamics and transport}
\label{sec:thermo}

Throughout this section we work on the AdS-Schwarzschild member $\alpha=0$, where the boundary thermodynamics is well defined, and we hold the physical source $\lambda_{-}$ fixed, so that $\eps\propto\rh$ by \eqref{eq:lambdaminus}.
We use the metric functions $u$, $v$ constructed in Section~\ref{sec:alpha0}.

\subsection{Equation of state}
\label{sec:eos}

\paragraph{Temperature and entropy.}
In the areal gauge \eqref{eq:arealmetric} the horizon stays at $r=\rh$ and its area density is uncorrected, so $s=\rh^{2}/4G$ exactly.
With the boundary time normalized by $u(\infty)=0$, the temperature is
\begin{equation}
T=T_{0}\big(1+\tau\eps^{2}\big),\qquad T_{0}=\frac{3\rh}{4\pi l^{2}},\qquad
\tau=\tfrac12\big[u(\rh)+v(\rh)\big],
\label{eq:dT}
\end{equation}
where $\tau$ is a pure number, independent of $\rh$ and $l$, because $u$ and $v$ depend on $r$ only through $x$.
Here $v(\rh)=-\tfrac13$ exactly, \eqref{eq:vh}.

\paragraph{The first law.}
At fixed $\lambda_{-}$ the $\Order{\eps^{2}}$ equation of state follows from \eqref{eq:dT} alone.
Since $\eps^{2}\propto\rh^{2}\propto s$, the speed of sound is
\begin{equation}
c_{s}^{2}=\Big(\frac{\partial\ln T}{\partial\ln s}\Big)_{\lambda_{-}}=\frac12+\tau\,\eps^{2}+\Order{\eps^{3}},
\label{eq:cs2first}
\end{equation}
and integrating $\dd p=s\,\dd T$, with $s(T)$ obtained by inverting \eqref{eq:dT}, gives
\begin{equation}
p=p_{0}\Big(1-\frac65\,\tau\,\eps^{2}\Big),\qquad p_{0}=\frac{\rh^{3}}{16\pi Gl^{2}},
\label{eq:pfirst}
\end{equation}
up to a temperature-independent constant, which vanishes in perturbation theory by dimensional analysis.

\paragraph{Free energy from holographic renormalization.}
Independently, the $\Order{\eps^{2}}$ free-energy density at fixed $T$ follows from the quadratic on-shell action of $\phim$.
Because $\sqrt{-g}\,g^{rr}=(r^{4}-2ml^{2}r)/l^{2}$ and $\psi_{1}$ are both independent of $\alpha$, the finite part of the boundary flux of the linear profile is independent of the hair,
\begin{equation}
\Big[\sqrt{-g}\,g^{rr}\,\psi_{1}\,\psi_{1}'\Big]^{\mathrm{fin}}_{r\to\infty}=-\frac{3\,C_{s}C_{v}\,\rh^{3}}{l^{2}},
\end{equation}
giving the bare bulk contribution $F^{(2)}_{\rm bulk}=-3C_{s}C_{v}\,\rh^{3}\eps^{2}/(32\pi Gl^{2})$, where the density is with respect to the $(x_{1},x_{2})$ plane.
The source branch of $\psi_{1}$ is non-normalizable, and the standard counterterm quadratic in $\phim$, whose coefficient is fixed by the leading divergence \cite{deHaro,BFS,Skenderis}, contributes a finite part of the same form, converting the factor $d=3$ into $2\Delta_{-}-d=5$:
\begin{equation}
F^{(2)}=-\frac{(2\Delta_{-}-d)\,C_{s}C_{v}\,\rh^{3}}{32\pi G\,l^{2}}\,\eps^{2}
=-\frac{3\sqrt3}{64\pi^{2}}\,\frac{\rh^{3}}{G\,l^{2}}\,\eps^{2}.
\label{eq:dF}
\end{equation}
This is the familiar statement $F^{(2)}=-\tfrac12\lambda_{-}\langle\Om\rangle$, with the response normalized by $(2\Delta_{-}-d)$ times the coefficient of the normalizable mode.

\paragraph{Consistency and closed form.}
Equating $-F^{(2)}$ with the $\Order{\eps^{2}}$ part of \eqref{eq:pfirst} fixes the temperature shift analytically,
\begin{equation}
\boxed{\;\tau=-\frac{25}{12}\,C_{s}C_{v}=-\frac{5\sqrt3}{8\pi}\approx-0.3446\;},
\qquad\text{i.e.}\qquad
u(\rh)=\frac13-\frac{5\sqrt3}{4\pi}.
\label{eq:tauclosed}
\end{equation}
The direct numerical integration of the metric equation \eqref{eq:u} reproduces this value of $u(\rh)$ to fifteen significant digits (Section~\ref{sec:alpha0} and Appendix~\ref{app:numerics}).
The agreement is a nontrivial check of the whole construction: it relates the purely geometric horizon data $u(\rh)$, $v(\rh)$ of the back-reacted metric to the purely asymptotic connection data $C_{s}C_{v}$ of the linear profile, through the first law and holographic renormalization.
It also shows that the $\Order{\eps^{2}}$ equilibrium thermodynamics of this irrelevant deformation is free of scheme ambiguities: a finite counterterm at this order would have to be a local functional of $\lambda_{-}$ and the flat boundary metric, of dimension $d=3$ and quadratic in a coupling of dimension $-1$, and no such functional exists. The scheme dependence of Section~\ref{sec:second} enters only at $\Order{\eps^{3}}$.

\paragraph{Equation of state.}
Collecting,
\begin{equation}
\boxed{\;
\frac{T-T_{0}}{T_{0}}\bigg|_{s}=-\frac{5\sqrt3}{8\pi}\,\eps^{2},\qquad
\frac{p^{(2)}}{p_{0}}=\frac{3\sqrt3}{4\pi}\,\eps^{2},\qquad
c_{s}^{2}=\frac12-\frac{5\sqrt3}{8\pi}\,\eps^{2},\qquad
\eps=\frac{4\pi\lambda_{-}T}{3C_{s}}\;}
\label{eq:EOS}
\end{equation}
up to $\Order{\eps^{3}}$ corrections.
At fixed entropy the deformation lowers the temperature, and at fixed temperature it raises the pressure.
Since $p^{(2)}\propto T^{5}$, the energy density and trace follow as $\edens^{(2)}=4p^{(2)}$ and $\langle T^{\mu}{}_{\mu}\rangle^{(2)}=2p^{(2)}-\edens^{(2)}=-2p^{(2)}$, consistent with the dilatation Ward identity $\langle T^{\mu}{}_{\mu}\rangle=(d-\Delta_{-})\lambda_{-}\langle\Om\rangle$.

\subsection{Transport coefficients}
\label{sec:transport}

The back-reaction also determines the leading first-order transport coefficients of the dual fluid.

\paragraph{Shear viscosity.}
The transverse-traceless graviton $h^{x_{1}}{}_{x_{2}}$ obeys, at zero spatial momentum, a minimally coupled massless wave equation, and the membrane-paradigm argument \cite{BuchelLiu,IqbalLiu} gives $\eta=h(\rh)^{2}/16\pi G$ for any horizon area density.
With $s=h(\rh)^{2}/4G$,
\begin{equation}
\frac{\eta}{s}=\frac{1}{4\pi}
\end{equation}
holds for any $\alpha$ and to all orders in $\eps$, as for any two-derivative Einstein-scalar theory \cite{KSS}.

\paragraph{Bulk viscosity.}
The Eling-Oz relation \cite{ElingOz}, derived from the null focusing equation on the horizon, expresses the bulk viscosity of a gravity-scalar system without chemical potential through the horizon values $\phi_{i}^{H}\equiv\phi_{i}(\rh)$ of the scalars along the equilibrium family,
\begin{equation}
\frac{\zeta}{\eta}=\sum_{i}\Big(s\frac{\dd\phi_{i}^{H}}{\dd s}\Big)^{2}=\sum_{i}c_{s}^{4}\,T^{2}\Big(\frac{\dd\phi_{i}^{H}}{\dd T}\Big)^{2}.
\label{eq:EO}
\end{equation}
With the normalizations $\psi_{1}(\rh)=1$ and $\psi_{2}(\rh)=0$, $\phim^{H}=\eps$ through second order, while $\pp^{H}=\pp^{(0)}(\rh)+\eps^{2}\Phi_{2}(\rh)$ enters \eqref{eq:EO} only at $\Order{\eps^{4}}$.
At fixed physical source $\lambda_{-}$, however, the hair parameter is \emph{not} constant.
Since $\lambda_{-}=\eps C_{s}l^{2}/\rh+\Order{\eps^{2}}$ by \eqref{eq:lambdaminus}, holding $\lambda_{-}$ fixed requires $\eps\propto\rh$: a larger brane needs a larger horizon value to produce the same boundary coefficient.
With $s=\rh^{2}/4G$ exact in the areal gauge, $\eps\propto s^{1/2}$, so that $s\,\dd\eps/\dd s=\tfrac12\eps\,[1+\Order{\eps}]$ and
\begin{equation}
\boxed{\;\frac{\zeta}{\eta}=\frac14\,\eps^{2}+\Order{\eps^{3}}\;}.
\label{eq:zeta}
\end{equation}
Although \eqref{eq:EO} was originally motivated in the high-temperature and adiabatic regimes, it has since been found to agree with the bulk viscosity extracted from the Kubo formula \cite{GubserPufuRocha} in the cascading and $\mathcal{N}=2^{*}$ plasmas, including their near-conformal and near-critical regimes \cite{BuchelEO}, and in non-trivial holographic RG-flow models \cite{BuchelGursoyKiritsis}.
The result \eqref{eq:zeta} is the leading near-conformal behavior: with $\phi^{H}\propto\lambda T^{\Delta-d}$ one has $s\,\dd\phi^{H}/\dd s=(\Delta-d)\phi^{H}/(d-1)$, so that the bulk viscosity is quadratic in the source and in $\Delta-d$, in line with the general near-conformal analyses of \cite{BuchelBulk,Yarom}.
Since \eqref{eq:EO} involves only horizon data and the entropy, \eqref{eq:zeta} is scheme independent.
Holding the hair parameter $\eps$ fixed instead would give $\zeta=0$ trivially, since the fixed-$(\alpha,\eps)$ solutions form a single orbit of the scaling symmetry of Section~\ref{sec:bg}.
That orbit is not a thermal path of the theory with fixed $\lambda_{-}$: under the scaling $\lambda_{-}\to\lambda_{-}/\Omega$ and $\mathcal{V}_{-}\to\Omega^{4}\mathcal{V}_{-}$, so that fixed $\eps$ is selected by a different, scale-invariant boundary condition, $\mathcal{V}_{-}\propto\lambda_{-}^{-4}$, i.e.\ a multi-trace functional $\mathcal{W}\propto\lambda_{-}^{-3}$ \cite{Witten,Papadimitriou}.
This is the analogue for $\Om$ of the fixed-$\hat k$ theory of \cite{YunExact}, and to the order considered here it yields conformal equilibrium thermodynamics.
Unlike there, however, the analogy is not exact: the running condensate of Section~\ref{sec:second} introduces the scale $\mu$, under which $\mu\,\partial_{\mu}\big(\mathcal{V}_{-}\lambda_{-}^{4}\big)=3\mathsf{A}\,C_{s}^{6}\eps^{6}\neq0$, so that the condition is scale invariant, and the fixed-$\eps$ free energy proportional to $T^{3}$, only up to the logarithm, which first enters at $\Order{\eps^{3}}$.
Since the source branch of $\phim$ is not normalizable (Section~\ref{sec:scheme}), such a theory is established at most in equilibrium, and the Dirichlet condition at fixed $\lambda_{-}$ adopted here is the standard one; the bulk viscosity, a Kubo response computed at fixed sources, is then \eqref{eq:zeta}.

\paragraph{Sound attenuation.}
In $d=3$ the sound mode $\omega=c_{s}k-i\Gamma_{s}k^{2}$ has attenuation constant $\Gamma_{s}=(\eta+\zeta)/[2(\edens+p)]$.
Using the exact Euler relation $\edens+p=Ts$ and $\eta/s=1/4\pi$,
\begin{equation}
4\pi T\,\Gamma_{s}=\frac12\Big(1+\frac{\zeta}{\eta}\Big)=\frac12+\frac18\,\eps^{2}+\Order{\eps^{3}},
\label{eq:Gamma}
\end{equation}
likewise scheme independent.

\paragraph{Sound speed and the bulk-viscosity bound.}
The sound speed is given by \eqref{eq:EOS}.
Combining it with \eqref{eq:zeta},
\begin{equation}
\frac{\zeta/\eta}{2\big(\frac12-c_{s}^{2}\big)}=\frac{\pi}{5\sqrt3}\approx0.36 ,
\end{equation}
so the lower bound $\zeta/\eta\geq2(\tfrac{1}{d-1}-c_{s}^{2})$ proposed in \cite{BuchelBulk} is not satisfied.
That bound was motivated by relevant deformations and is known not to be universal \cite{BuchelViolation}; the present model provides a simple, analytically controlled example in which an irrelevant deformation violates it at leading order.

\paragraph{Summary.}
To $\Order{\eps^{2}}$ on AdS-Schwarzschild, $\eta/s=1/4\pi$ exactly, while $\zeta/\eta$, $4\pi T\Gamma_{s}$ and $c_{s}^{2}$ depart from their conformal values by closed-form, scheme-independent amounts.
For $\alpha>0$, $\eta/s=1/4\pi$ continues to hold.
The bulk viscosity, by contrast, requires the equilibrium family at fixed boundary data, including the data that fix the $\pp$-hair itself.
In the Dirichlet theory of \cite{YunExact}, $\alpha$ varies along the thermal path and the single-scalar background already has $\zeta\neq0$; combining this with $\lambda_{-}$ requires the dictionary for $\lambda_{-}$ at $\alpha>0$ noted in Section~\ref{sec:bg}, and we do not quote $\zeta$ for $\alpha>0$.
The relaxation rate of $\Om$ itself, the lowest quasinormal frequency of $\phim$ at zero momentum, is not of this type and is left for future work.

\section{A third scalar: the \texorpdfstring{$\mathbb{Z}_{2}$}{Z2}-odd field \texorpdfstring{$\chi$}{chi}}
\label{sec:chi}

The clean solvability of the $\phim$ analysis rests on two accidents of the truncation: the vanishing mixed mass $V_{,+-}|_{0}=0$, and the exact cancellation of the hair function $S$ between the wave operator and the scalar mass.
In this and the next section we ask how much of this structure survives the addition of the third scalar of the ABJM sector.

\subsection{The three-scalar model and nested truncations}
\label{sec:chimodel}

The action \eqref{eq:action} extends to \cite{Gauntlett,BakYun}
\begin{equation}
I=\frac{1}{16\pi G}\int\dd^{4}x\,\sqrt{-g}\Big[\,R-\tfrac12(\partial\pp)^{2}
-\tfrac12(\partial\phim)^{2}-\tfrac32\,K(\pp,\phim)\,(\partial\chi)^{2}
-V(\pp,\phim,\chi)\Big],
\label{eq:action3}
\end{equation}
with the field-dependent kinetic normalization and potential
\begin{align}
K(\pp,\phim)&=e^{-\frac{2\pp}{\sqrt7}-\frac{4\phim}{\sqrt{21}}},
\label{eq:K}\\[2pt]
V&=-\frac{12}{l^{2}}e^{-\frac{3\pp}{\sqrt7}+\frac{\phim}{\sqrt{21}}}
+\frac{3}{2l^{2}}e^{-\frac{3\pp}{\sqrt7}+\frac{8\phim}{\sqrt{21}}}
+\frac{9}{2l^{2}}e^{-\sqrt7\,\pp}\big(1+\chi^{2}\big)^{2}
+\frac{6}{l^{2}}\,\chi^{2}\,e^{-\frac{5\pp}{\sqrt7}+\frac{4\phim}{\sqrt{21}}}.
\label{eq:V3}
\end{align}
Setting $\chi=0$ returns \eqref{eq:action} and \eqref{eq:V}.
The stress tensor acquires the term $3K[\partial_{\mu}\chi\partial_{\nu}\chi-\tfrac12g_{\mu\nu}(\partial\chi)^{2}]$, and the scalar equations become
\begin{equation}
\Box\pp=V_{,+}+\tfrac32 K_{,+}(\partial\chi)^{2},\qquad
\Box\phim=V_{,-}+\tfrac32 K_{,-}(\partial\chi)^{2},\qquad
3\,\nabla_{\mu}\!\big(K\nabla^{\mu}\chi\big)=V_{,\chi},
\label{eq:eom3}
\end{equation}
with $K_{,a}\equiv\partial K/\partial\varphi_{a}$.
Henceforth $\cdot|_{0}$ denotes evaluation at $\phim=\chi=0$.

\paragraph{A $\mathbb{Z}_{2}$-protected truncation.}
Both $K$ and $V$ are even under $\chi\to-\chi$, and $K$ is independent of $\chi$.
Hence $V_{,\chi}|_{\chi=0}=0$ identically, and $\chi=0$ solves the $\chi$ equation for any configuration of the dilatons.
This is a symmetry statement, sharper than the tuned cancellation \eqref{eq:notadpole} that protects the $\phim$ truncation.
The same parity forces $V_{,+\chi}|_{0}=V_{,-\chi}|_{0}=0$, so that $\chi$ does not mix with either dilaton at linear order.
The kinetic weight on the truncation is $K|_{0}=S^{-1}$, and the local mass of $\chi$ is
\begin{equation}
m_{\chi}^{2}\equiv\frac{V_{,\chi\chi}|_{0}}{3K|_{0}}=\frac{4}{l^{2}}\,S^{-3/2}+\frac{6}{l^{2}}\,S^{-5/2}.
\label{eq:mchi}
\end{equation}
At the AdS vacuum $m_{\chi}^{2}l^{2}=10$, and the dual operator has dimension
\begin{equation}
\Delta_{\chi}=\tfrac32+\sqrt{\tfrac94+10}=5=d+2 ,
\end{equation}
one unit more irrelevant than $\Om$, and again of integer dimension.

\paragraph{Nested truncations.}
Setting a scalar to zero identically turns its own equation of motion into a constraint on the others, and for \eqref{eq:V3} these constraints are strong.
We consider static configurations, for which $(\partial\chi)^{2}=g^{rr}\chi'^{2}\geq0$ outside the horizon.

\emph{(i) $\pp\equiv0$ forces $\phim\equiv0$ and $\chi\equiv0$.}
With $\Box\pp=0$ the first of \eqref{eq:eom3} becomes, writing $\theta\equiv\phim/\sqrt{21}$,
\begin{equation}
0=\frac{Q(\theta)}{\sqrt7\,l^{2}}
-\frac{9\sqrt7}{l^{2}}\chi^{2}-\frac{9\sqrt7}{2l^{2}}\chi^{4}
-\frac{30}{\sqrt7\,l^{2}}\,e^{4\theta}\chi^{2}
-\frac{3}{\sqrt7}\,e^{-4\theta}(\partial\chi)^{2},\qquad
Q(\theta)=36e^{\theta}-\tfrac92e^{8\theta}-\tfrac{63}{2}.
\label{eq:nestA}
\end{equation}
Since $Q'(\theta)=36e^{\theta}(1-e^{7\theta})$, the only critical point of $Q$ is its maximum $Q(0)=0$, so $Q\leq0$; every other term is manifestly nonpositive.
All terms must therefore vanish separately, pointwise, with no boundary condition needed.

\emph{(ii) $\phim\equiv0$ forces $\chi\equiv0$ on any black hole with a regular, nondegenerate horizon.}
The second of \eqref{eq:eom3} becomes
\begin{equation}
V_{,-}+\tfrac32K_{,-}(\partial\chi)^{2}\Big|_{\phim=0}
=\frac{6}{\sqrt{21}}\,S^{-1}\Big[\frac{4}{l^{2}}\,S^{-3/2}\chi^{2}
-(\partial\chi)^{2}\Big]=0 .
\label{eq:nestB}
\end{equation}
At the horizon $g^{rr}=0$ while $\chi'$ is finite, so $\chi(\rh)=0$.
Away from it, $|\chi'|=\tfrac2l S^{-3/4}|\chi|/\sqrt{g^{rr}}$, whose coefficient is integrable at a simple zero of $g^{rr}$, and Gr\"onwall's inequality gives $\chi\equiv0$.
On the background \eqref{eq:bg}, where $g^{rr}=fS^{-3/2}$, the factor $S$ drops out and \eqref{eq:nestB} reads $f\chi'^{2}=4\chi^{2}/l^{2}$ for every $\alpha$, with general solution $\chi=c\,\chi_{c}^{\pm1}$, $\chi_{c}=\exp\!\big[\tfrac2l\int^{r}\dd r'/\sqrt{f}\,\big]$; since $\chi_{c}'\propto(r-\rh)^{-1/2}$, regularity requires $c=0$.

The truncations thus nest: $\chi=0$ is protected by $\mathbb{Z}_{2}$, $\phim=0$ is consistent only inside it, and $\pp=0$ only when both other scalars vanish.
Read in reverse, a nonzero $\phim$ necessarily drives $\pp$, and a nonzero $\chi$ necessarily drives both dilatons.
These are the exact, nonlinear counterparts of the second-order sources: the vertex $V_{,+--}|_{0,S=1}=Q''(0)/(21\sqrt7\,l^{2})\neq0$ that generates $\Phi_{2}$ in \eqref{eq:Phi2}, the nonpositive $\chi$-terms of \eqref{eq:nestA} that generate the $\Order{\delta^{2}}$ correction to $\pp$, and the bracket of \eqref{eq:nestB} that generates the $\Order{\delta^{2}}$ profile of $\phim$ (Section~\ref{sec:deltadelta}).

\subsection{Linear \texorpdfstring{$\chi$}{chi} hair on AdS-Schwarzschild}
\label{sec:chilin}

We switch on $\chi$ with amplitude $\delta$.
By the $\mathbb{Z}_{2}$ symmetry, $\chi$ carries only odd powers of $\delta$, while the even fields receive corrections at even orders; together with the $\phim$ hair of the previous sections,
\begin{align}
\chi &=\delta\,\chi_{1}+\eps\delta\,\chi_{11}+\delta^{3}\chi_{3}+\cdots,
\label{eq:jointchi}\\
\phim &=\eps\,\psi_{1}+\eps^{2}\,\psi_{2}+\delta^{2}\,\psi_{2}^{(\delta\delta)}+\cdots,
\label{eq:jointphim}\\
\{\pp,g_{\mu\nu}\}&=\{\pp^{(0)},g^{(0)}_{\mu\nu}\}
+\eps^{2}\{\Phi_{2},g^{(2)}_{\mu\nu}\}+\delta^{2}\{\Phi_{2}^{\chi},g^{(2)}_{\chi\,\mu\nu}\}+\cdots .
\label{eq:jointmetric}
\end{align}
We work to total degree two in $(\eps,\delta)$; no hierarchy between the amplitudes is needed at this order.
At $\Order{\delta}$ the metric and the dilatons are frozen, and a static profile $\chi_{1}(r)$ satisfies, using $K|_{0}=S^{-1}$ and $V_{,\chi\chi}|_{0}$ from Appendix~\ref{app:V},
\begin{equation}
\boxed{\;\frac{\dd}{\dd r}\!\Big(\frac{f\,r^{2}}{S}\,\chi_{1}'\Big)
=\frac{r^{2}}{l^{2}}\Big(\frac{6}{S^{2}}+\frac{4}{S}\Big)\chi_{1}\;}\,.
\label{eq:chir}
\end{equation}
In contrast to \eqref{eq:linr}, the hair function does \emph{not} cancel: the two mass contributions carry different powers of $S$, and the kinetic weight $S^{-1}$ does not match $\sqrt{-g_{0}}$.

On AdS-Schwarzschild ($S\equiv1$), \eqref{eq:chir} collapses to the hypergeometric equation
\begin{equation}
x(x-1)\,\ddot\chi_{1}+(2x-1)\,\dot\chi_{1}-\tfrac{10}{9}\,\chi_{1}=0,
\qquad (a,b;c)=\big(\tfrac53,-\tfrac23;1\big),
\label{eq:hypchi}
\end{equation}
to be compared with $(\tfrac43,-\tfrac13;1)$ for $\phim$.
The solution regular at the horizon is $\chi_{1}=\Fhyp(\tfrac53,-\tfrac23;1;1-x)=1+\tfrac{10}{9}t+\Order{t^{2}}$, and by the connection formulas \cite{DLMF} it is the superposition of a source branch $C_{s}^{\chi}(x-1)^{2/3}\Fhyp(-\tfrac23,-\tfrac23;-\tfrac43;\tfrac1{1-x})$ and a response branch $C_{v}^{\chi}(x-1)^{-5/3}\Fhyp(\tfrac53,\tfrac53;\tfrac{10}{3};\tfrac1{1-x})$, with
\begin{equation}
\boxed{\;C_{s}^{\chi}=\frac{\Gamma(\tfrac13)}{\Gamma(\tfrac23)^{2}}\approx1.4610,
\qquad
C_{v}^{\chi}=-\frac{3}{7}\,\frac{\Gamma(\tfrac23)}{\Gamma(\tfrac13)^{2}}\approx-0.0809\;},
\qquad C_{s}^{\chi}C_{v}^{\chi}=-\frac{3\sqrt3}{14\pi}.
\label{eq:CsCvchi}
\end{equation}
The fall-offs $x^{2/3}\sim r^{2}$ and $x^{-5/3}\sim r^{-5}$ are the source and response of a dimension-five operator, so that $\delta\propto\lambda_{\chi}T^{2}$ at fixed physical source $\lambda_{\chi}$.
Since $C_{s}^{\chi}\neq0$, the horizon-regular mode carries a nonzero source: there is no spontaneous $\chi$ hair, and switching on $\chi$ breaks $\mathbb{Z}_{2}$ explicitly.
A genuinely spontaneous condensate would require an additional scale, as in the holographic superconductors driven by a chemical potential \cite{Gubser,HHH}, including their top-down realizations in M-theory \cite{GSW} and in the charged sector of ABJM \cite{BakKimYun}.

\subsection{The hairy background: a Heun equation with an apparent singularity}
\label{sec:chiheun}

For $\alpha>0$, substituting $S=\beta_{1}x+\beta_{0}$ with $\beta_{1}=2\sinh\alpha$, $\beta_{0}=e^{-\alpha}$ into \eqref{eq:chir} gives
\begin{equation}
x(x-1)\,S\,\ddot\chi_{1}+\big(\beta_{1}\,x^{2}+2\beta_{0}\,x-\beta_{0}\big)\dot\chi_{1}
=\tfrac19\big(4S+6\big)\chi_{1}.
\label{eq:heun}
\end{equation}
The leading coefficient $x(x-1)S$ is cubic, so \eqref{eq:heun} is a Fuchsian equation with four regular singular points,
\begin{equation}
\big\{0,\;1,\;x_{*},\;\infty\big\},\qquad
x_{*}=-\frac{\beta_{0}}{\beta_{1}}=-\frac{e^{-\alpha}}{2\sinh\alpha}<0 ,
\end{equation}
i.e.\ a Heun equation \cite{Ronveaux}.
The extra singular point sits at $S=0$, where $\pp^{(0)}\to-\infty$, outside the physical interval.

\paragraph{Apparency and its protection.}
The point $x_{*}$ is an \emph{apparent} singularity.
Its indicial exponents are $\{0,2\}$, and the would-be logarithm in the smaller-exponent solution is obstructed only by the combination $-e^{2\alpha}+2e^{\alpha}\sinh\alpha+1$, which vanishes by the identity \eqref{eq:identity} that makes the background a solution.
The full Riemann scheme is
\begin{equation}
\left\{\begin{matrix}
x=0 & x=1 & x_{*} & \infty\\[2pt]
0 & 0 & 0 & -\tfrac23\\
0 & 0 & 2 & \tfrac23
\end{matrix}\right\},
\label{eq:riemann}
\end{equation}
whose exponents sum to $2$, as Fuchs' relation requires.
Because the monodromy around $x_{*}$ is trivial, the projective monodromy of \eqref{eq:heun} is generated by $\{0,1,\infty\}$ alone, as for a hypergeometric equation, and the single accessory parameter is fixed by the apparency condition rather than free.
Equation \eqref{eq:heun} admits no elementary solution $S^{a}x^{b}(x-1)^{c}$, since the exponents at $x_{*}$ and at $\infty$ are incompatible; and for generic $\alpha$ it is not reducible to a single Gauss function by a change of variable and prefactor, since such reductions require the four singular points to sit at special cross-ratios \cite{Maier}, whereas $x_{*}$ moves continuously with $\alpha$.

\paragraph{Closed form.}
The apparency nevertheless makes a reduction to Gauss functions explicit.
In the standard Heun normalization\footnote{We write the Heun parameters with hats to distinguish them from the amplitudes and the hair parameter: $y''+\big(\tfrac{\hat\gamma}{x}+\tfrac{\hat\delta}{x-1}+\tfrac{\hat\varepsilon}{x-x_{*}}\big)y'+\tfrac{\hat\alpha\hat\beta x-\hat q}{x(x-1)(x-x_{*})}\,y=0$.} \eqref{eq:heun} has $\hat\gamma=\hat\delta=1$, $\hat\varepsilon=-1$, and $\hat\alpha\hat\beta=-\tfrac49$ with $\hat\alpha+\hat\beta=0$, so that the exponent $1-\hat\varepsilon=2$ at $x_{*}$ is a nonzero integer.
Heun equations of this type admit generalized hypergeometric solutions when, in addition, the accessory parameter $\hat q$ obeys a polynomial condition \cite{LetessierValentWimp,Takemura,Ishkhanyan}, which for $\hat\varepsilon=-1$ is quadratic in $\hat q$ and is equivalent to the apparency of $x_{*}$ \cite{Takemura}.
For \eqref{eq:heun} this condition reduces, once more, to the identity \eqref{eq:identity}, and is therefore satisfied for every $\alpha$.
The regular solution normalized to $\chi_{1}(1)=1$ is
\begin{equation}
\boxed{\;\chi_{1}(x)={}_{3}F_{2}\!\Big(\tfrac23,-\tfrac23,1+\tfrac23e^{\alpha};\,
1,\tfrac23e^{\alpha};\,1-x\Big)
=\Big[1+\tfrac32e^{-\alpha}(x-1)\,\partial_{x}\Big]
\Fhyp\!\big(\tfrac23,-\tfrac23;1;1-x\big)\;}\,,
\label{eq:chi3F2}
\end{equation}
a first-order differential operator acting on a single Gauss function.
That Gauss function $\mathcal{G}$ solves the three-point equation $x(x-1)\ddot{\mathcal{G}}+x\dot{\mathcal{G}}-\tfrac49\mathcal{G}=0$, with exponents $\{0,1\}$, $\{0,0\}$ and $\{-\tfrac23,\tfrac23\}$ at $x=0,1,\infty$: these are the singular points of \eqref{eq:riemann} without $x_{*}$, the unit shift at $x=0$ compensating, through Fuchs' relation, the exponent $2$ carried by the apparent point.
As $\alpha\to0$ the lower parameter $\tfrac23e^{\alpha}$ cancels the upper parameter $\tfrac23$, and \eqref{eq:chi3F2} collapses to $\Fhyp(\tfrac53,-\tfrac23;1;1-x)$, as it must.
We have verified \eqref{eq:chi3F2} against \eqref{eq:heun} numerically (Appendix~\ref{app:numerics}).

\paragraph{Non-AdS asymptotics.}
As $x\to\infty$ with $\beta_{1}\neq0$, \eqref{eq:heun} reduces to $x^{2}\ddot\chi_{1}+x\dot\chi_{1}-\tfrac49\chi_{1}=0$, with exponents $\pm\tfrac23$, i.e.\ $\chi_{1}\sim r^{\pm2}$ rather than the dimension-five fall-offs of Section~\ref{sec:chilin}; this asymptotic equation is exactly the underlying Gauss equation of \eqref{eq:chi3F2}.
This is in marked contrast with $\phim$, whose fall-offs are insensitive to the hair.
Applying the Gauss connection formulas to the underlying Gauss function, and using that $(x-1)\partial_{x}$ acts on the leading powers $(x-1)^{\pm2/3}$ as $\pm\tfrac23$,
\begin{equation}
\chi_{1}=C_{+}(\alpha)\,(x-1)^{2/3}\big[1+\Order{x^{-1}}\big]+C_{-}(\alpha)\,(x-1)^{-2/3}\big[1+\Order{x^{-1}}\big],
\label{eq:Cpm}
\end{equation}
where each bracket denotes the complete branch, and
\begin{equation}
C_{+}(\alpha)=\frac{\Gamma(\tfrac13)}{2\Gamma(\tfrac23)^{2}}\big(1+e^{-\alpha}\big),\qquad
C_{-}(\alpha)=-\frac{3\,\Gamma(\tfrac23)}{2\,\Gamma(\tfrac13)^{2}}\big(1-e^{-\alpha}\big).
\label{eq:Cpmcoef}
\end{equation}
As $\alpha\to0$, $C_{+}\to C_{s}^{\chi}$ while $C_{-}\to0$: the limit of the asymptotics is singular, the dimension-five scaling being recovered only inside a boundary layer $x\lesssim(2\sinh\alpha)^{-1}$.
Assigning $C_{\pm}$ the roles of source and response requires the adapted renormalization mentioned in Section~\ref{sec:bg}.
Since $\sqrt{-g_{0}}\,K|_{0}\,g_{0}^{rr}=fr^{2}/S\propto r$ at large $r$, the $r^{+2}$ branch has divergent kinetic norm while the $r^{-2}$ branch is normalizable; as $C_{+}(\alpha)\neq0$, the regular $\chi$ hair is non-normalizable for every $\alpha$, and the absence of spontaneous $\chi$ hair extends to the whole family, independently of how $C_{\pm}$ are eventually interpreted.

\section{The third scalar at second order: cross-sourcing without running}
\label{sec:chisecond}

The order counting of \eqref{eq:jointchi}--\eqref{eq:jointmetric} is transparent.
At $\Order{\eps^{2}}$ nothing changes: the results of Sections~\ref{sec:second}--\ref{sec:thermo} hold verbatim, since the $\chi$-dependent terms of \eqref{eq:K} and \eqref{eq:V3} do not contribute there.
At $\Order{\delta^{2}}$ the $\chi$ condensate sources $\phim$ (through $V_{,-\chi\chi}$ and $K_{,-}$), $\pp$ and the metric.
Since $V_{,-\chi}$ is odd in $\chi$ and $K$ is $\chi$-independent, there is \emph{no} $\Order{\eps\delta}$ source for $\phim$, $\pp$ or the metric: the entire $\Order{\eps\delta}$ physics lives in the $\chi$ equation, as a correction $\chi_{11}$ driven by the $\phim$ hair.
We carry out the resonance analysis on the AdS-Schwarzschild member $\alpha=0$, where every profile is hypergeometric and the dictionary is standard.

\subsection{Resonances and a one-step identity}
\label{sec:onestep}

On AdS-Schwarzschild all linear operators we meet belong to the family
\begin{equation}
\Lop_{\sigma}\equiv x(x-1)\,\partial_{x}^{2}+(2x-1)\,\partial_{x}-\sigma(\sigma+1),
\label{eq:Lsigma}
\end{equation}
with $\sigma=\tfrac13$ for $\phim$ ($\Lop_{1/3}=\Lop$ of \eqref{eq:hyp}), $\sigma=\tfrac23$ for $\chi$, and $\sigma=1$ for $\pp$.
The dual dimension is $\Delta=d(\sigma+1)$, and the source and response fall-offs are $x^{\sigma}$ and $x^{-\sigma-1}$.
On powers,
\begin{equation}
\Lop_{\sigma}\,x^{q}=P_{\sigma}(q)\,x^{q}-q^{2}\,x^{q-1},\qquad
P_{\sigma}(q)=(q-\sigma)(q+\sigma+1),
\label{eq:Lpowers}
\end{equation}
so asymptotic expansions descend in integer steps of $x^{-1}\propto m$, i.e.\ in powers of the energy density.
Every second-order scalar equation has the form $\Lop_{\sigma_{3}}[w]=\mathcal{S}$, with a forcing bilinear in two linear profiles $u$, $v$ solving $\Lop_{\sigma_{1}}[u]=0$ and $\Lop_{\sigma_{2}}[v]=0$.
Two general facts control whether such a forcing produces a logarithm.

\emph{Kinetic couplings are reducible.}
Writing $c_{i}\equiv\sigma_{i}(\sigma_{i}+1)$, a direct computation gives
\begin{equation}
\Lop_{\sigma_{3}}[uv]=2x(x-1)\,\dot u\,\dot v+(c_{1}+c_{2}-c_{3})\,uv ,
\label{eq:kinred}
\end{equation}
of which \eqref{eq:Lpsi1sq} is the special case $u=v=\psi_{1}$.
A forcing $\kappa_{1}uv+\kappa_{2}\,x(x-1)\dot u\dot v$, of the type produced by a potential vertex and a field-dependent kinetic term, therefore equals $\big[\kappa_{1}-\tfrac12\kappa_{2}(c_{1}+c_{2}-c_{3})\big]uv$ plus $\tfrac12\kappa_{2}\Lop_{\sigma_{3}}[uv]$, and the latter is solved by the logarithm-free function $\tfrac12\kappa_{2}uv$.
Logarithms can thus come only from the product $uv$, with an effective coupling.

\emph{Resonances.}
A forcing tower $\sum_{n\ge0}J_{p-n}x^{p-n}$ is matched by $\sum_{n}w_{p-n}x^{p-n}$ with $P_{\sigma_{3}}(q)w_{q}-(q+1)^{2}w_{q+1}=J_{q}$.
If the tower reaches a zero $q_{*}$ of $P_{\sigma_{3}}$, a term $b\,x^{q_{*}}\ln x$ is required, with
\begin{equation}
b\,P_{\sigma_{3}}'(q_{*})=R,\qquad R\equiv J_{q_{*}}+(q_{*}+1)^{2}\,w_{q_{*}+1},
\label{eq:Rdef}
\end{equation}
so that the effective resonant forcing $R$ collects every power of the tower at or above the resonance, not only the resonant coefficient itself \cite{Skenderis,vanReesA}.
A logarithm at the response exponent makes the condensate run; one at the source exponent signals a logarithmic mixing into the source.

\emph{The one-step identity.}
A single branch $u\simeq x^{s}(1+u_{1}x^{-1}+\cdots)$ of $\Lop_{\sigma}$, with $s\in\{\sigma,-\sigma-1\}$, has $u_{1}=-\tfrac{s}{2}$, as follows from \eqref{eq:Lpowers} with $P_{\sigma}(s-1)=-2s$.
Let $u$ and $v$ be branches with leading exponents $s_{1}$ and $s_{2}$, and suppose the product $uv$, of leading power $p=s_{1}+s_{2}$, forces an operator $\Lop_{\sigma_{3}}$ that is resonant exactly one step below, at $q_{*}=p-1$, i.e.\ $c_{3}=p(p-1)$.
Then $P_{\sigma_{3}}(p)=2p$, $w_{p}=1/(2p)$, and
\begin{equation}
R=\Big(-\frac{s_{1}}{2}-\frac{s_{2}}{2}\Big)+\frac{p^{2}}{P_{\sigma_{3}}(p)}
=-\frac{p}{2}+\frac{p^{2}}{2p}=0 .
\label{eq:onestep}
\end{equation}
\emph{A product of two linear profiles never produces a logarithm at a resonance one step below its leading power}, whatever the couplings.
The $\phim$ self-coupling of Section~\ref{sec:second} escapes \eqref{eq:onestep} because its resonance lies two steps down.
We now show that both cross-couplings meet only one-step resonances, or none.

\subsection{Order \texorpdfstring{$\delta^{2}$}{delta squared}: \texorpdfstring{$\chi$}{chi} sources \texorpdfstring{$\phim$}{phi-}}
\label{sec:deltadelta}

By the nested-truncation statement (ii), a nonzero $\chi$ necessarily sources $\phim$.
The cross-vertex does not vanish on the truncation,
\begin{equation}
V_{,-}\big|_{\phim=0}=\frac{24}{\sqrt{21}\,l^{2}}\,\chi^{2}\,e^{-5\pp/\sqrt7}
\ \ (\text{exactly}),\qquad
V_{,-\chi\chi}\big|_{0,S=1}=\frac{48}{\sqrt{21}\,l^{2}} ,
\label{eq:vertex}
\end{equation}
and the kinetic term contributes a further source through $K_{,-}=-\tfrac{4}{\sqrt{21}}K$.
On AdS-Schwarzschild the induced profile obeys
\begin{equation}
\begin{gathered}
\big(\Box-m_{-}^{2}\big)\psi_{2}^{(\delta\delta)}
=\tfrac12 V_{,-\chi\chi}\big|_{0}\,\chi_{1}^{2}
+\tfrac32 K_{,-}\big|_{0}\,g^{rr}\,(\chi_{1}')^{2},\\
\text{i.e.}\qquad
\Lop_{1/3}\big[\psi_{2}^{(\delta\delta)}\big]=\frac{2}{3\sqrt{21}}\Big[4\chi_{1}^{2}-9\,x(x-1)\,\dot\chi_{1}^{2}\Big],
\end{gathered}
\label{eq:psi2eom}
\end{equation}
where we used $\Box-m_{-}^{2}=\tfrac{9}{l^{2}}\Lop_{1/3}$ and $g^{rr}(\chi_{1}')^{2}=\tfrac{9}{l^{2}}x(x-1)\dot\chi_{1}^{2}$.
The bracket is that of \eqref{eq:nestB}.

\paragraph{Neither resonance is excited.}
The branches of $\chi_{1}$ have leading powers $x^{2/3}$ and $x^{-5/3}$, so the forcing lies in the towers $x^{4/3-n}$, $x^{-1-n}$ and $x^{-10/3-n}$.
None contains the response power $x^{-4/3}$ of $\Lop_{1/3}$: the vacuum resonance is out of reach kinematically, and the $\chi$ hair cannot make $\langle\Om\rangle$ run.
The source power $x^{1/3}$ lies exactly one step below the leading power of the first tower, and \eqref{eq:onestep}, with $s_{1}=s_{2}=\tfrac23$ and $c_{3}=\tfrac49=\tfrac43\cdot\tfrac13$, shows that it produces no logarithm either, for any values of $V_{,-\chi\chi}$ and $K_{,-}$.
Hence
\begin{equation}
\boxed{\;\mathsf{A}_{-\chi\chi}\;\equiv\;
\frac{\mu\,\partial_{\mu}\mathcal{V}_{-}}{\lambda_{\chi}^{2}}\;=\;0\;}\,,
\end{equation}
despite the nonzero vertex \eqref{eq:vertex}: the $\chi$ condensate induces neither a running of $\langle\Om\rangle$ nor a logarithmic mixing into $\lambda_{-}$.
Explicitly, evaluating \eqref{eq:Rdef} with the exact branch expansions gives $R=0$ for the potential and kinetic halves of \eqref{eq:psi2eom} separately, whereas a pure $x^{4/3}$ forcing gives $R=\tfrac23\neq0$.

\paragraph{Saturation of the constraint.}
The nested-truncation constraint adds a stronger, separate statement.
The forcing in \eqref{eq:psi2eom} vanishes identically when $\chi$ obeys $3\sqrt{x(x-1)}\,\partial_{x}\chi=\pm2\chi$, solved by $\chi_{c}^{\pm1}$ with $\chi_{c}=(\sqrt{x}+\sqrt{x-1})^{4/3}$.
The coefficient $4$ in the bracket equals $(\Delta_{\chi}-d)^{2}$: the ratio $V_{,-\chi\chi}|_{0,S=1}=-3K_{,-}|_{0,S=1}(\Delta_{\chi}-d)^{2}/l^{2}$ makes the pure-AdS source mode $\chi\propto r^{\Delta_{\chi}-d}$ saturate the bracket, so that in pure AdS a constant $\chi$ source induces no $\phim$ at all at this order.
On AdS-Schwarzschild $\chi_{c}$ is no longer a solution, but only barely, $\Lop_{2/3}[\chi_{c}]=\tfrac1{12}x^{-2}\chi_{c}[1+\Order{x^{-1}}]$, and the source branch of $\chi_{1}$, $x^{2/3}(1-\tfrac13x^{-1}+\tfrac1{18}x^{-2}+\cdots)$, agrees with $2^{-4/3}\chi_{c}=x^{2/3}(1-\tfrac13x^{-1}-\tfrac5{72}x^{-2}+\cdots)$ through next-to-leading order.
The forcing therefore loses its two leading orders: the potential and kinetic contributions at $x^{1/3}$ are equal and opposite,
\begin{equation}
\mathcal{S}^{\rm pot}\big|_{x^{1/3}}=-\,\mathcal{S}^{\rm kin}\big|_{x^{1/3}}=-\frac{16\sqrt{21}}{189}\,(C_{s}^{\chi})^{2},
\qquad
\mathcal{S}\big|_{x^{4/3}}=\mathcal{S}\big|_{x^{1/3}}=0,
\label{eq:cancel}
\end{equation}
and $\psi_{2}^{(\delta\delta)}$ is forced only from the nonresonant power $x^{-2/3}$ downward (Appendix~\ref{app:source}).
For a different ratio of couplings the leading orders would survive and $\psi_{2}^{(\delta\delta)}$ would grow faster at large $x$, but by \eqref{eq:onestep} still no logarithm would appear.

\paragraph{Vacuum anomaly.}
These statements concern the homogeneous thermal state.
They are distinct from the vacuum conformal anomaly of $\langle\Om\Oc\Oc\rangle$ \cite{BzowskiMcFaddenSkenderis}, which occurs at $\Delta_{-}+2\Delta_{\chi}-2d=8=2k$ with $k=4$ and multiplies eight derivatives of the sources; its computation requires the momentum-dependent Witten diagram \cite{FMMR} and is not addressed here.

\subsection{Order \texorpdfstring{$\eps\delta$}{epsilon delta}: \texorpdfstring{$\phim$}{phi-} does not run the \texorpdfstring{$\chi$}{chi} condensate}
\label{sec:epsdelta}

At $\Order{\eps\delta}$ the $\chi$ equation acquires a source from the interference of $\psi_{1}$ and $\chi_{1}$ through the same two couplings $V_{,-\chi\chi}$ and $K_{,-}$.
On AdS-Schwarzschild, $3\Box-V_{,\chi\chi}|_{0,S=1}=\tfrac{27}{l^{2}}\Lop_{2/3}$, and simplifying on-shell with $\partial_{x}[x(x-1)\dot\chi_{1}]=\tfrac{10}{9}\chi_{1}$,
\begin{equation}
\boxed{\;\Lop_{2/3}[\chi_{11}]
=\frac{56}{9\sqrt{21}}\,\psi_{1}\chi_{1}
+\frac{4}{\sqrt{21}}\,x(x-1)\,\dot\psi_{1}\dot\chi_{1}\;}\,,
\label{eq:chi11}
\end{equation}
the first term coming from the potential and the second from the kinetic coupling.
A logarithm in $\langle\Oc\rangle$ requires a tower of the forcing to reach the response power $x^{-5/3}$ of $\Lop_{2/3}$.

The forcing organizes into four towers, with leading powers $x^{1}$, $x^{-2/3}$, $x^{-4/3}$ and $x^{-3}$.
The response power $x^{-5/3}$ sits in the $x^{-2/3}$ tower, which is supplied \emph{solely} by the product of the $\phim$ response branch and the $\chi$ source branch, with leading coefficient $C_{v}C_{s}^{\chi}$.
The two coefficients that control the resonance are
\begin{equation}
J_{-2/3}=C_{v}C_{s}^{\chi}\,\frac{8}{3\sqrt{21}},\qquad
J_{-5/3}=C_{v}C_{s}^{\chi}\,\frac{8}{9\sqrt{21}} ,
\label{eq:Tcoeffs}
\end{equation}
and \eqref{eq:Rdef}, with $w_{-2/3}=-\tfrac34J_{-2/3}$ and $P_{2/3}'(-\tfrac53)=-\tfrac73$, gives
\begin{equation}
b_{\chi}=-\frac{3J_{-5/3}-J_{-2/3}}{7}=0 .
\label{eq:bslope}
\end{equation}
The $\phim$ hair thus induces \emph{no} running of the $\chi$ condensate at $\Order{\eps\delta}$.
The mechanism is again \eqref{eq:onestep}: the resonance $x^{-5/3}$ lies one step below the leading power of the product of branches with $s_{1}=-\tfrac43$ and $s_{2}=\tfrac23$, and $c_{3}=\tfrac{10}{9}=(-\tfrac53)(-\tfrac23)$.
Accordingly $b_{\chi}=0$ for any ratio of $V_{,-\chi\chi}$ to $K_{,-}$, as a direct evaluation of \eqref{eq:Rdef} with arbitrary coefficients confirms.
The source power $x^{2/3}$ is likewise unforced: the only tower congruent to it modulo integers, $x^{-4/3-n}$, begins below it, so its coefficient is a choice of normalization.

\subsection{The condensates through second order}
\label{sec:combinedcond}

Assembling the orders, and quoting the normalizable coefficients in units of $(\rh/l^{2})^{\Delta}$, the two condensates through total degree two read
\begin{align}
\frac{\mathcal{V}_{-}}{(\rh/l^{2})^{4}}
&=\eps\,C_{v}
+\eps^{2}\Big[a_{-4/3}'+3\mathsf{A}\,C_{s}^{2}\ln\frac{\mu l^{2}}{\rh}\Big]
+\delta^{2}\,a_{-4/3}^{(\delta\delta)}+\cdots,
\label{eq:combinedOm}\\
\frac{\mathcal{V}_{\chi}}{(\rh/l^{2})^{5}}
&=\delta\,C_{v}^{\chi}+\eps\delta\,a_{-5/3}^{(\eps\delta)}+\cdots,
\label{eq:combinedOchi}
\end{align}
with $a^{(\delta\delta)}_{-4/3}$ and $a^{(\eps\delta)}_{-5/3}$ finite constants and \emph{no} logarithm beyond the single $\Order{\eps^{2}}$ term of \eqref{eq:OmVEV}.
The running of each condensate is fixed entirely by its own self-coupling, the cubic vertex $V_{,---}$ for $\Om$ (and none for $\Oc$, whose cubic self-coupling is forbidden by $\mathbb{Z}_{2}$), while every cross-coupling contributes only a finite shift.
The two sectors thus decouple, as far as running is concerned, even while sharing one geometry.

\subsection{\texorpdfstring{$\chi$}{chi}-induced geometry, temperature and transport}
\label{sec:chigeom}

\paragraph{Back-reaction.}
On AdS-Schwarzschild the $\Order{\delta^{2}}$ metric corrections add to those of $\phim$, since there is no $\Order{\eps\delta}$ metric source; we write $g_{tt}\supset-f\delta^{2}u_{\chi}$ and $g^{rr}\supset f\delta^{2}v_{\chi}$ in the areal gauge \eqref{eq:arealmetric}.
The general equations \eqref{eq:v}--\eqref{eq:u} apply with $k=3$ (from the normalization $\tfrac32K$) and $M^{2}l^{2}=V_{,\chi\chi}|_{0,S=1}\,l^{2}=30$, and the closed form \eqref{eq:vclosed} gives
\begin{equation}
\boxed{\;v_{\chi}(x)=-\tfrac94\,x\,\chi_{1}\dot\chi_{1}\;},\qquad
v_{\chi}(\rh)=-\tfrac94\cdot1\cdot\tfrac{10}{9}=-\tfrac52 ,
\label{eq:vchiclosed}
\end{equation}
the exact, purely geometric counterpart of $v(\rh)=-\tfrac13$.
At large $x$, $v_{\chi}\simeq-\tfrac32(C_{s}^{\chi})^{2}x^{4/3}$, so the perturbative geometry breaks down at $r\sim\rh\,\delta^{-1/2}$, where $\lambda_{\chi}E^{2}\sim1$.
The dilaton correction obeys
\begin{equation}
\Lop_{1}[\Phi_{2}^{\chi}]=-\frac{31}{3\sqrt7}\,\chi_{1}^{2}-\frac{3}{\sqrt7}\,x(x-1)\,\dot\chi_{1}^{2},
\label{eq:Phi2chi}
\end{equation}
sourced by $V_{,+\chi\chi}$ and $K_{,+}$ in accordance with the nested truncation (i), and is integrated with the homogeneous solutions \eqref{eq:Phihom}.

\paragraph{UV sensitivity of the temperature.}
Unlike $u$, the function $u_{\chi}$ does not settle at large $r$: the leading growth of the $(rr)$ equation does not cancel, and
\begin{equation}
u_{\chi}'(r)=\frac{9\,(C_{s}^{\chi})^{2}}{4\,\rh}+\Order{r^{-3}},
\label{eq:uchiasym}
\end{equation}
so $u_{\chi}$ diverges linearly and the canonical normalization $u_{\chi}(\infty)=0$ cannot be imposed.
The corresponding term in $g_{tt}$ grows faster than the background and cannot be removed by a constant rescaling of $t$.
Its removal at a cutoff $r=\Lambda$ corresponds to a divergence $\propto\lambda_{\chi}^{2}\Lambda\langle T_{tt}\rangle$, and the associated finite counterterm, with an arbitrary scale $\mu$, shifts the temperature by a term $\propto\lambda_{\chi}^{2}\mu T^{3}$.
The $\chi$-induced temperature shift is therefore UV-sensitive: its scheme-independent content is exhausted by the horizon datum $v_{\chi}(\rh)$, while the full shift requires the counterterms that a $\Delta=d+2$ deformation needs at second order in its source \cite{vanReesA,vanReesB}, including those associated with the growing metric modes.
The route that fixed the $\phim$ thermodynamics in Section~\ref{sec:thermo} would determine the $\Order{\delta^{2}}$ equation of state once this renormalization is completed.
The pattern is the expected one: the more irrelevant the operator, the more of its thermodynamic response is boundary data rather than horizon geometry.

\paragraph{Transport.}
The shear viscosity ratio $\eta/s=1/4\pi$ is unaffected.
To the order considered, the Eling-Oz formula \eqref{eq:EO} applies to $\chi$ after a constant rescaling: on the horizon of the $\alpha=0$ background, $K=1-\tfrac{4}{\sqrt{21}}\,\eps+\Order{\eps^{2},\delta^{2}}$, so that $\hat\chi=\sqrt3\,\chi$ is canonically normalized up to field-dependent corrections that contribute to $\zeta/\eta$ only at $\Order{\eps\delta^{2}}$, and \eqref{eq:EO} holds for the canonical scalars $(\pp,\phim,\hat\chi)$ \cite{ElingOz,HoyosKimOz}.
With $\chi^{H}=\delta\propto\rh^{2}\propto s$ at fixed $\lambda_{\chi}$, so that $s\,\dd\delta/\dd s=\delta$, \eqref{eq:zeta} and \eqref{eq:Gamma} become
\begin{equation}
\boxed{\;\frac{\zeta}{\eta}=\frac14\,\eps^{2}+3\,\delta^{2},\qquad
4\pi T\,\Gamma_{s}=\frac12+\frac18\,\eps^{2}+\frac32\,\delta^{2}\;}
\label{eq:zetachi}
\end{equation}
up to third-order corrections.
Since both involve only horizon data and the entropy, they are insensitive to the UV issue above.
The $\Order{\delta^{2}}$ correction to $c_{s}^{2}$, by contrast, awaits the renormalization just discussed.

\section{Discussion}
\label{sec:disc}

We have constructed, to second order in the amplitude, a planar AdS$_{4}$ black hole carrying hair for both dilatons of the ABJM truncation, and extended the analysis to its third, $\mathbb{Z}_{2}$-odd scalar.
The construction is essentially analytic.
The linearized $\phim$ profile is an exact hypergeometric function, independent of the hair, whose horizon-regular branch always carries a source, so the second scalar enters only as an irrelevant deformation together with an induced condensate (Section~\ref{sec:scheme}).
At second order the irrelevance of $\Om$ forces a two-step resonance whose coefficient $\mathsf{A}=\sqrt{21}/70=l^{2}V_{,---}|_{0,S=1}/120$ encodes a logarithmically running condensate (Section~\ref{sec:second}); the running is state dependent, multiplying $\lambda_{-}^{2}\langle T_{tt}\rangle^{2}$.
The back-reaction is closed form: the $g_{rr}$ correction is algebraic in the linear profile for any minimally coupled scalar, and for arbitrary hair a single P\"oschl-Teller pair controls the metric and $\pp$ response, which excites only one eigendirection and locks $\Phi_{2}=\tfrac{6}{\sqrt7}H_{2}$ (Section~\ref{sec:backreact}).
On AdS-Schwarzschild, the first law and the renormalized on-shell action agree and yield the analytic equation of state \eqref{eq:EOS}; the agreement ties the horizon data $u(\rh)$, $v(\rh)$ to the asymptotic connection data $C_{s}C_{v}$ and shows that the $\Order{\eps^{2}}$ thermodynamics is scheme independent (Section~\ref{sec:thermo}).
The shear viscosity ratio stays at $1/4\pi$, while the bulk viscosity, sound attenuation and sound speed depart from their conformal values by closed-form amounts, in violation of the bulk-viscosity bound of \cite{BuchelBulk}.
The main results are collected in Table~\ref{tab:summary}.

\begin{table}[t]
\centering
\renewcommand{\arraystretch}{1.35}
\begin{tabular}{lll}
\hline
Quantity & Result & Equation\\
\hline
Linear $\phim$ profile (all $\alpha$) & $\Fhyp(\tfrac43,-\tfrac13;1;1-x)$, $\;C_{s}C_{v}=\tfrac{3\sqrt3}{10\pi}$ & \eqref{eq:psi1}, \eqref{eq:CsCv}\\
Log coefficient of $\psi_{2}$ & $\mathsf{A}\,C_{s}^{2}$, $\;\mathsf{A}=\tfrac{\sqrt{21}}{70}=\tfrac{l^{2}}{120}V_{,---}|_{0,S=1}$ & \eqref{eq:psi2bndry}, \eqref{eq:anomcubic}\\
$g_{rr}$ corrections ($\alpha=0$) & $v=-\tfrac{3k}{4}x\varphi\dot\varphi$, $\;v(\rh)=-\tfrac13$, $\;v_{\chi}(\rh)=-\tfrac52$ & \eqref{eq:vclosed}, \eqref{eq:vchiclosed}\\
Back-reaction (all $\alpha$) & $\Phi_{2}=\tfrac{6}{\sqrt7}H_{2}$, $\;\bm{f}_{2}\cdot\bm{S}=0$ & \eqref{eq:sigma}\\
Temperature at fixed $s$ & $T/T_{0}-1=-\tfrac{5\sqrt3}{8\pi}\eps^{2}$ & \eqref{eq:tauclosed}\\
Sound speed & $c_{s}^{2}=\tfrac12-\tfrac{5\sqrt3}{8\pi}\eps^{2}$ & \eqref{eq:EOS}\\
Viscosities & $\eta/s=\tfrac{1}{4\pi}$, $\;\zeta/\eta=\tfrac14\eps^{2}+3\delta^{2}$ & \eqref{eq:zeta}, \eqref{eq:zetachi}\\
Linear $\chi$ profile ($\alpha>0$) & Clausen function ${}_{3}F_{2}$, apparent singularity & \eqref{eq:chi3F2}\\
Cross-couplings & $\mathsf{A}_{-\chi\chi}=0$, $\;b_{\chi}=0$ for any couplings & \eqref{eq:onestep}, \eqref{eq:bslope}\\
\hline
\end{tabular}
\caption{Summary of the main closed-form results. Statements about sources, condensates and thermodynamics refer to $\alpha=0$.}
\label{tab:summary}
\end{table}

The third scalar preserves some of this structure and breaks the rest in an instructive way (Sections~\ref{sec:chi} and \ref{sec:chisecond}).
Its truncation is protected by symmetry, and on AdS-Schwarzschild its linear analysis is hypergeometric.
On the hairy background the hair ceases to decouple and the linear problem becomes a Heun equation; its fourth singularity is apparent, protected by the very identity that underlies the background, and its regular solution is a Clausen function with closed-form hairy connection coefficients.
At second order the $\chi$ condensate necessarily sources $\phim$, since the truncations nest, and the $\phim$ hair corrects $\chi$; yet neither cross-coupling makes either condensate run, and the two sectors decouple, as far as running is concerned, while sharing one geometry.
The $\chi$ back-reaction mirrors the dilaton's closed form, $v_{\chi}(\rh)=-\tfrac52$, but the greater irrelevance of $\Oc$ pushes its temperature response into the ultraviolet.

\paragraph{Origin of the mutual non-renormalization.}
The absence of cross-logarithms is not a property of the couplings: on AdS-Schwarzschild both $\mathsf{A}_{-\chi\chi}=0$ and $b_{\chi}=0$ hold for any ratio of $V_{,-\chi\chi}$ to $K_{,-}$.
What they require is that each cross-coupling resonance lie exactly one step, one power of the energy density, below its forcing.
In terms of $\sigma=\Delta/d-1$, the $\Order{\delta^{2}}$ condition reads $2\sigma_{\chi}-1=\sigma_{-}$ and the $\Order{\eps\delta}$ condition reads $(\sigma_{\chi}-\sigma_{-}-1)-1=-\sigma_{\chi}-1$; both are equivalent to the single relation
\begin{equation}
2\Delta_{\chi}-\Delta_{-}=2d ,
\label{eq:dimrel}
\end{equation}
obeyed here ($10-4=6$).
The vacuum resonance of $\Om$ is out of reach at $\Order{\delta^{2}}$ because $[2(\Delta_{\chi}-d)+\Delta_{-}]/d=\tfrac83$ is not an integer, whereas for the $\phim$ self-coupling the analogous number $(3\Delta_{-}-2d)/d=2$ is an integer, so that resonance is met two steps down and the condensate runs.
The dimensions $\Delta_{-}=d+1$ and $\Delta_{\chi}=d+2$ belong to operators of the ABJM spectrum dual to the scalars of the truncation.
If their values are fixed by superconformal multiplet structure, as their integrality suggests, then \eqref{eq:dimrel}, and with it the mutual non-renormalization, would be a consequence of the representation theory of the ABJM fixed point rather than of the dynamics of the truncation.
The vacuum counterpart of the question, whether the conformal anomaly of $\langle\Om\Oc\Oc\rangle$ at $\Delta_{-}+2\Delta_{\chi}-2d=8$ vanishes as well, is natural to pose in the momentum-space framework of \cite{BzowskiMcFaddenSkenderis} and through the Osborn equation for irrelevant operators \cite{SchwimmerTheisen,Broccoli}; a vanishing with a symmetry origin, in the sense of \cite{PetkouSkenderis}, would test this expectation directly.

\paragraph{Further directions.}
A boundary-theory computation of $\mathsf{A}$ would complete the identification of the $\phim$ logarithm; the multi-trace reading suggested in Section~\ref{sec:second} indicates where to look.
The fixed ratio $6/\sqrt7$ between the metric and $\pp$ responses is a consequence of the constant matrix $\bm{M}_{A}$ and presumably reflects an underlying symmetry of the truncation that deserves to be made explicit.
Within the Dirichlet theory of \cite{YunExact}, the present results are the leading terms at $T\ll\Lambda$.
Since $\psi_{1}$ and $\psi_{2}$ are independent of $\alpha$, extending them to finite $T/\Lambda$ requires only the dictionary for $\lambda_{-}$ on the hyperscaling-violating background, an extension of the renormalization of \cite{YunExact,KST}; it would show how the $\phim$ deformation interacts with the $\pp$ flow at relative order $(T/\Lambda)^{3}$, and would make the source/response meaning of $C_{\pm}(\alpha)$ of \eqref{eq:Cpm} precise.
The Clausen form \eqref{eq:chi3F2} makes the corresponding second-order $\chi$ analysis at $\alpha>0$ feasible in principle.
Completing the $\Order{\lambda_{\chi}^{2}}$ renormalization would fix the $\chi$ temperature shift and sound speed, and the first-law check that worked for $\phim$ would then provide a sharp test.
The distinction between horizon and boundary data should also govern other probes: for the holographic entanglement entropy \cite{RT} of a large region the extensive part is fixed by the horizon, whose area density is uncorrected at second order in the areal gauge, whereas the subleading terms probe the near-boundary region; within the perturbative window one may also ask how entanglement $c$-functions \cite{MyersSinha,LiuMezei} respond to an irrelevant deformation, which leaves the ultraviolet fixed point.
Finally, the perturbative breakdown at $r\sim\rh/\eps$ and $r\sim\rh\,\delta^{-1/2}$, characteristic of irrelevant deformations, motivates a complete nonlinear treatment with an explicit ultraviolet completion, for which the present results provide a controlled infrared seed.

\section*{Acknowledgments}

The author is grateful to J.~Kim for helpful discussions.

\appendix

\section{Potential data on the truncation}
\label{app:V}

With $S=e^{2\pp/\sqrt7}$ and all quantities evaluated at $\phim=0$ (and $\chi=0$ for the three-scalar model), the derivatives of \eqref{eq:V} used in Sections~\ref{sec:model}--\ref{sec:thermo} are
\begin{align}
V_{,-}&=0, & V_{,+-}&=0, \\
m_{-}^{2}=V_{,--}&=\frac{4}{l^{2}}\,S^{-3/2}, &
V_{,---}&=\frac{12\sqrt{21}}{7\,l^{2}}\,S^{-3/2}, \\
V_{,+--}&=-\frac{12\sqrt7}{7\,l^{2}}\,S^{-3/2}, &
V_{,++}&=\frac{1}{l^{2}}\Big(-\tfrac{27}{2}\,S^{-3/2}+\tfrac{63}{2}\,S^{-7/2}\Big),\\
V_{,+}&=\frac{9\sqrt7}{2\,l^{2}}\,\big(S^{2}-1\big)S^{-7/2}. & &
\end{align}
In the gauge of Section~\ref{sec:generalalpha}, with $W_{0}=e^{2B_{0}}=\tfrac{l^{2}m^{2}}{9}\,\sinh^{3/2}(m\rho+\alpha)/\sinh^{7/2}(m\rho)$, these combine to give
\begin{equation}
W_{0}\,m_{-}^{2}=\frac{4m^{2}}{9}\csch^{2}(m\rho),\qquad
W_{0}\,V_{,+--}=-\frac{4\sqrt7\,m^{2}}{21}\csch^{2}(m\rho).
\end{equation}
For the three-scalar model \eqref{eq:K}--\eqref{eq:V3}, the additional data used in Sections~\ref{sec:chi} and \ref{sec:chisecond} are
\begin{align}
&V_{,\chi}=V_{,+\chi}=V_{,-\chi}=0,\qquad
V_{,\chi\chi}=\frac{18}{l^{2}}S^{-7/2}+\frac{12}{l^{2}}S^{-5/2},\qquad
V_{,-\chi\chi}=\frac{48}{\sqrt{21}\,l^{2}}S^{-5/2},\\
&V_{,+\chi\chi}=-\frac{18\sqrt7}{l^{2}}S^{-7/2}-\frac{60}{\sqrt7\,l^{2}}S^{-5/2},\qquad
K=S^{-1},\qquad K_{,+}=-\frac{2}{\sqrt7}\,S^{-1},\qquad K_{,-}=-\frac{4}{\sqrt{21}}\,S^{-1},
\end{align}
while the $\chi$-independent data above are unchanged.

\section{Asymptotic recursion and the logarithm}
\label{app:recursion}

In the variable $t=x-1$ the operator of \eqref{eq:hyp} acts on powers as
\begin{equation}
\Lop[t^{q}]=\Big(q+\tfrac43\Big)\Big(q-\tfrac13\Big)\,t^{q}+q^{2}\,t^{q-1},
\end{equation}
which is triangular in descending powers, with the source and response indices as the zeros of $P(q)=(q+\tfrac43)(q-\tfrac13)$.
From \eqref{eq:psi1bndry}, the source branch of $\psi_{1}$ is $C_{s}t^{1/3}(1+\tfrac16t^{-1}-\tfrac19t^{-2}+\cdots)$, so that
\begin{equation}
\psi_{1}^{2}=C_{s}^{2}\,t^{2/3}+\tfrac{1}{3}C_{s}^{2}\,t^{-1/3}+2C_{s}C_{v}\,t^{-1}-\tfrac{7}{36}C_{s}^{2}\,t^{-4/3}+\cdots .
\end{equation}
Matching order by order in \eqref{eq:psi2x} fixes
\begin{equation}
a_{2/3}=\frac{\sqrt{21}}{7}C_{s}^{2},\qquad a_{-1/3}=\frac{\sqrt{21}}{21}C_{s}^{2},\qquad a_{-1}=-\frac{9}{\sqrt{21}}C_{s}C_{v}.
\end{equation}
The index $q=\tfrac13$ is also a zero of $P$, but no power of the forcing is congruent to it, so its coefficient $A_{s}$ remains free; it is the normalization freedom discussed in Section~\ref{sec:second}.
At $q=-\tfrac43$ a pure power cannot balance the forcing.
Including $b\,t^{-4/3}\ln t$, for which $\Lop[t^{-4/3}\ln t]=-\tfrac53\,t^{-4/3}+\Order{t^{-7/3}\ln t}$, the balance
\begin{equation}
-\tfrac53\,b+\tfrac19\,a_{-1/3}=\frac{2}{\sqrt{21}}\Big(-\tfrac{7}{36}C_{s}^{2}\Big)
\end{equation}
yields
\begin{equation}
b=\frac{\sqrt{21}}{70}\,C_{s}^{2}=\mathsf{A}\,C_{s}^{2}.
\end{equation}
We have confirmed this value, together with $A_{s}\approx-0.590$ and $a_{-4/3}\approx0.207$, by fitting the numerically integrated regular solution.

\section{The \texorpdfstring{$\Order{\delta^{2}}$}{O(delta squared)} source and its resonant coefficients}
\label{app:source}

On AdS-Schwarzschild the horizon-regular $\chi_{1}$ has the boundary expansion, in $\varsigma\equiv x^{-1/3}$,
\begin{equation}
\chi_{1}=C_{s}^{\chi}\,\varsigma^{-2}\big(1-\tfrac13 \varsigma^{3}+\tfrac1{18}\varsigma^{6}+\cdots\big)
+C_{v}^{\chi}\,\varsigma^{5}\big(1+\tfrac56 \varsigma^{3}+\cdots\big),
\end{equation}
from which $\chi_{1}^{2}$ contains the powers $x^{4/3},x^{1/3},x^{-2/3},\dots$ from the source branch and $x^{-1},\dots$ from the cross term, but no $x^{-4/3}$.
The kinetic term $x(x-1)\dot\chi_{1}^{2}$ shares this structure.
Reading off the coefficients of the forcing in \eqref{eq:psi2eom}, $\mathcal{S}=\mathcal{S}^{\rm pot}+\mathcal{S}^{\rm kin}$ with $\mathcal{S}^{\rm pot}=\tfrac{8}{3\sqrt{21}}\chi_{1}^{2}$ and $\mathcal{S}^{\rm kin}=-\tfrac{6}{\sqrt{21}}x(x-1)\dot\chi_{1}^{2}$, at the powers $x^{4/3}$, $x^{1/3}$ and $x^{-4/3}$ gives \eqref{eq:cancel}: the response power is absent term by term, and the source power cancels between the two pieces.
The absence of a logarithm does not rely on this cancellation, by \eqref{eq:onestep}.
The forcing is nonzero at the neighboring nonresonant powers, for instance $\mathcal{S}|_{x^{-2/3}}=\tfrac{2\sqrt{21}}{21}(C_{s}^{\chi})^{2}$, confirming that the vanishing is specific to the resonances and not an overall triviality.

\section{Numerical checks}
\label{app:numerics}

All analytic statements in the main text that were not derived in closed form were checked numerically; we summarize the procedures here.
Hypergeometric and Clausen functions were evaluated with arbitrary-precision arithmetic (30 significant digits), and all numbers below are quoted to the precision at which they are stable under changes of the integration range and precision.

\paragraph{Temperature shift.}
The function $u$ was obtained by integrating \eqref{eq:u} with $\varphi=\psi_{1}$ and $v$ from \eqref{eq:vclosed}, in the variable $x$, by double-exponential (tanh-sinh) quadrature on $1<x<10^{15}$, with the analytic tail $\int^{\infty}u'\,\dd r$ added using $u'\simeq3C_{s}^{2}\rh/8r^{2}$.
The result, $u(\rh)=-0.355827785943907$, differs from the closed form $\tfrac13-\tfrac{5\sqrt3}{4\pi}$ by less than $10^{-15}$.
The same procedure for $\chi$ confirms that $u_{\chi}'(r)-9(C_{s}^{\chi})^{2}/(4\rh)$ decays as $r^{-3}$, with no $r^{-1}$ or $r^{-2}$ terms.

\paragraph{Clausen solution and connection coefficients.}
Inserting \eqref{eq:chi3F2} into \eqref{eq:heun} at $\alpha=0.3$ and $\alpha=1.2$ gives residuals below $10^{-29}$ at all sample points, and the two forms in \eqref{eq:chi3F2} agree to the same accuracy.
The coefficients \eqref{eq:Cpmcoef} were checked by applying the operator of \eqref{eq:chi3F2} separately to the two Gauss branches of the underlying function and comparing the leading coefficients at $x=10^{3}$ and $10^{6}$; the agreement improves as $x^{-1}$, as expected from the subleading terms of the branches.

\paragraph{Second-order $\phim$ profile.}
Equation \eqref{eq:psi2x} was integrated from $t=10^{-3}$, with initial data from the Frobenius series \eqref{eq:psi2reg}, using an eighth-order Runge-Kutta method with relative tolerance $10^{-13}$.
The asymptotic form \eqref{eq:psi2bndry} was then fitted by least squares on $t\in[50,2\times10^{4}]$, keeping the forced coefficients fixed and fitting $A_{s}$, $a_{-4/3}$ and the coefficients of $t^{-5/3},t^{-2},t^{-7/3},t^{-7/3}\ln t,t^{-8/3},t^{-3}$.
This gives $A_{s}=-0.58961$ and $a_{-4/3}=0.2071$; leaving the logarithmic coefficient free returns $0.08391$, to be compared with $\mathsf{A}C_{s}^{2}=0.08390$.

\paragraph{Field equations and resonance coefficients.}
All potential data of Appendix~\ref{app:V}, the second-order equations \eqref{eq:psi2x}, \eqref{eq:Phi2}, \eqref{eq:psi2eom}, \eqref{eq:chi11} and \eqref{eq:Phi2chi}, and the matrix decomposition \eqref{eq:Mdecomp} were rederived symbolically from \eqref{eq:action3}.
The logarithmic coefficients $\mathsf{A}C_{s}^{2}$, $\mathsf{A}_{-\chi\chi}=0$ and $b_{\chi}=0$ were recomputed by solving the recursion \eqref{eq:Rdef} with the exact branch series, the last two also for arbitrary potential and kinetic couplings.

\paragraph{The $\Order{\delta^{2}}$ source.}
The large-$x$ expansion of the forcing in \eqref{eq:psi2eom}, evaluated directly from the hypergeometric $\chi_{1}$, reproduces the vanishing of the $x^{4/3}$ and $x^{1/3}$ coefficients and the value of the $x^{-2/3}$ coefficient quoted in Appendix~\ref{app:source}.
The expansions of $\chi_{c}$ and of the source branch of $\chi_{1}$ quoted in Section~\ref{sec:deltadelta}, as well as $\Lop_{2/3}[\chi_{c}]=\tfrac1{12}x^{-2}\chi_{c}[1+\Order{x^{-1}}]$, were verified symbolically.


\end{document}